\documentclass[12pt,preprint]{aastex}
\usepackage{amsmath, amssymb}

\slugcomment{Aug. 01 2009 version}

\shorttitle{Emission from a Young Protostellar Object}
\shortauthors{Yamada et al.}

\begin{document}

\title{Emission from a Young Protostellar Object I. Signatures of 
Young Embedded Outflows}

\author{Masako Yamada}
\affil{Institute of Astronomy and Astrophysics, Academia Sinica, P. O. Box 23-141, Taipei, 
 10617, Taiwan, R.O.C.}
\email{masako@asiaa.sinica.edu.tw}

\author{Masahiro N. Machida\altaffilmark{1}}
\affil{Department of Physics, Graduate School of Science, Kyoto University,
Sakyo-ku, Kyoto, 606-8502}

\author{Shu-ichiro Inutsuka}
\affil{Department of Physics, Nagoya University, Furo-cho, Chikusa-ku
Nagoya, 464-8602}
\and

\author{Kohji Tomisaka\altaffilmark{2}}
\affil{Division of Theoretical Astronomy and Center for Computational Astronomy, National Astronomical
  Observatory of Japan, Osawa, Mitaka, Tokyo, 181-8588}


\altaffiltext{1}{present address: Division of Theoretical Astronomy and Center for Computational
Astronomy, National Astronomical Observatory of Japan, Osawa, Mitaka, Tokyo, 181-8588}
\altaffiltext{2}{also at School of Physical Sciences, Graduate University for Advanced 
Studies (SOKENDAI).}

\begin{abstract}
We examine emission from a young protostellar object (YPO) with three-dimensional
ideal MHD simulations and three-dimensional non-local thermodynamic equilibrium (non-LTE) line
transfer calculations, and show the first results.
To calculate the emission field, we employed a snapshot result of an MHD simulation 
having young bipolar outflows and 
a dense protostellar disk (a young circumstellar disk) embedded in an infalling envelope.
Synthesized line emission of two molecular species (CO and SiO) show that
subthermally excited SiO lines as a high density tracer can provide a better 
probe of the complex velocity field of 
a YPO, compared to fully thermalized CO lines.
In a YPO at the earliest stage when the outflows are still embedded in the 
collapsing envelope, infall, rotation and outflow motions have similar speeds.
We find that the combined velocity field of these components introduces a great complexity in the 
line emissions through varying optical thickness and emissivity, such as 
asymmetric double-horn profiles.  
We show that the rotation of the outflows, 
one of the features that characterizes an outflow driven by magneto-centrifugal
forces, appears clearly in velocity channel maps and intensity-weighted mean velocity 
(first moment of velocity) maps.
The somewhat irregular morphology of the line emission at this youngest stage is dissimilar to 
a more evolved object such as young Class 0.
High angular resolution observation by e.g., 
the Atacama Large Millimeter/submillimeter Array (ALMA) telescope can reveal these
features.
Our results demonstrate a powerful potential of the synthesized emission of the three-dimensional 
line transfer to probe the velocity field embedded in the envelope, and further analysis will be able to determine the precise velocity field to assess the dynamics in the young protostellar object to 
gain a better understanding of star formation. 
\end{abstract}


\section{Introduction}

The formation of stars begins with the collapse of a parent dense molecular core.
As gravitational collapse of a molecular core proceeds, bipolar outflows are expected to
form in relation to accretion and rotation of the parent core.
Observations of the embedded, young Class 0 objects revealed that some of them
exhibit infall and rotational motions in emission profiles as well as bipolar outflows
(see e.g., \citealp{belloche2002} for IRAM 04191; \citealp{bourke2005} for L1041-IRS; 
\citealp{matthews2006} for Bernerd 1-c).
The velocity structure in the stage prior to Class 0 objects,  before the formation
of a protostar, is expected to represent the dynamics in the earliest phase of star formation.
Therefore, the observational studies of the very young stage of star formation, 
on the way from a prestellar (starless) core to a Class 0 object, are quite important in 
understanding the star formation processes.
We further can expect a hint of a launching mechanism of molecular outflows, 
almost free from the propagation effect,  in the earliest stage of star formation.

Before a protostar begins to shine, temperature of the star-forming object without
a heating source is as quite low as 10\,K.
Major observational ways are thus molecular emission lines in the millimeter and submillimeter bands.
After a protostar forms, the young object at the earliest stage of star formation 
(hereafter ``a young protostellar object" or a YPO for simplicity) has: 1) a vast amount of  
envelope matter, 2) complex velocity structure consisting of infall, rotation, and outflow motions, and
3) a complicated optical thickness structure due to the complex velocity field in the millimeter and
submillimeter bands.
As a result, it is not an easy task to correctly interpret the complex velocity field imprinted 
in the emergent line emission.
Furthermore, current observational facilities do not achieve sufficient angular 
resolution to resolve the varying velocity field within the molecular core.
However, forthcoming high angular resolution observations provided by the 
Atacama Large Millimeter/submillimeter Array (ALMA) will be able to reveal 
these internal structures of an evolving core, and a significant progress in understanding the 
star formation process is expected.
In addition, the launch of the {\em Herschel} satellite will enable observations 
with high frequency resolution in the submillimeter (and shorter wavelength) band.
For these new observational facilities, detailed modelling and understanding 
of the radiation field of young star forming objects is of crucial importance.
In this regard, prestellar cores have been studied with one-dimensional hydrodynamic model and 
radiative transfer calculations \citep[e.g.,][]{pav2008,tsamis2008}. 
Here, we employ a theoretical model constructed with a three-dimensional, high resolution numerical 
simulation of the formation process of a protostar \citep{machida2008}. 
By combining it with three-dimensional, 
non-local thermodynamic equilibrium (non-LTE) line transfer calculations,  we 
examine a characteristic feature in the radiation field of a YPO.
In this article, we show the first results in formats common to radio observers.
Feasibility evaluation of these synthesized emissions by the ALMA telescope is another 
interesting problem, but we do not treat it in detail in this article except for simple 
estimates of the necessary exposure time.

The outline of this article is as follows.
In \S2, we describe the models and the set-ups of the magnetohydrodynamic 
and radiative transfer simulations.
The results of continuum and molecular line emissions are presented in the same formats
used in radio observations in \S3. 
We make comments with regard to observations with current and future 
observational facilities and discuss them in \S4. 
Summary and conclusions are presented in \S5.

\section{Models and Simulations}

\subsection{Magnetohydrodynamic Modelling} \label{sect:hydro}

Evolution from a parent dense molecular core to a protostar proceeds with a several
characteristic evolutionary stages \citep[see e.g.,][and their Figure 2b]{masunaga2000}.
At the onset of collapse, cooling by dust emission is so effective within a molecular core 
that it contracts nearly isothermally with an initial low temperature.
Gravitational contraction proceeds and the central density increases, until the core 
becomes optically thick to the dust thermal emission.
At this stage the cooling rate is less efficient and the collapse becomes adiabatic
at the density of $n\sim 10^{11}$ cm$^{-3}$.
At this point an adiabatic core of a radius  
$R\sim1-10$ AU and a mass $\sim 0.01 M_\sun$ surrounded by the accretion 
shock is formed at the center.
This core is called ``the first core" \citep{larson1969}.
The kinetic temperature, $T_\mathrm{kin}$, increases as the first core contracts adiabatically.
When $T_\mathrm{kin}$ exceeds $\sim$ 2000 K, hydrogen molecules begin to dissociate.
Since dissociation is an endothermic process, the equation of state becomes soft again,
and it induces dynamical contraction of the central region at $n\gtrsim 10^{16}$
cm$^{-3}$ (second collapse). 
Once all the hydrogen molecules are dissociated, the equation of state becomes 
adiabatic again, and eventually a second core (protostar) is formed.

Considering this picture, we employed the MHD modelling 
of \citet{machida2008} as one among those most plausible and based on detailed calculations.
\citet{machida2008} investigated the evolution of a star-forming object
from a parent dense molecular core
to a protostar by three-dimensional resistive MHD simulations.
They showed that the magnetic force working in an evolving core can naturally drive 
a slow and wide opening angle flow of $\sim$ 5 km s$^{-1}$ and a fast and 
collimated flow of $\sim$ 30 km s$^{-1}$ from the first and second cores,
respectively.
In their scenario, these two flows are expected to increase their 
velocities and evolve to a molecular 
bipolar outflow ($v\gtrsim 10$ km s$^{-1}$) and an optical jet ($v\gtrsim 100$
km s$^{-1}$) as observed in more evolved young stellar objects (YSOs).
In this article, we follow their scenario and simply refer to the low-velocity flow formed
in their calculations as an ``outflow".
Observational study of protostellar objects and outflows in their earliest evolutionary stage
is indeed quite important for a comprehensive understanding of the star formation 
process as well as for the theoretical modelling.
Since the characteristic structures of the youngest star-forming objects are small in size, 
they will be a milestone target of special interest of the ALMA telescope.

\citet{machida2008} showed that ideal and resistive MHD modelings do not 
introduce a significant difference in the onset of molecular outflows around the first core.
In this study, we calculated the cloud collapse based on the ideal MHD equations:
\begin{mathletters}
\begin{eqnarray}
  \frac{\partial\rho}{\partial t} + \nabla\cdot (\rho \boldsymbol{v}) &=& 0,  \\
  \rho \frac{\partial \boldsymbol{v}}{\partial t} + \rho (\boldsymbol{v}\cdot\nabla)\boldsymbol{v}
    &=& -\nabla P -\frac{1}{4\pi}\boldsymbol{B}\times(\nabla \times\boldsymbol{B}) -\rho\nabla\phi, \\
  \frac{\partial\boldsymbol{B}}{\partial t} &=& \nabla\times (\boldsymbol{v}\times \boldsymbol{B}), \\
  \nabla^2\phi &=& 4\pi G\rho, 
\end{eqnarray}
\end{mathletters}
where $\phi$ denotes gravitational potential, and 
other symbols are used for usual meanings.

In order to calculate a model in a wide dynamic range, a nested grid method is adopted
\citep[see for details][]{machida2005}.
In this scheme, a rectangular grid of each level has the same number of grid points ($128^2\times 64$),
and cell width $h(l)$ varies with the grid level $l$.
The width of a cell $h(l)$ decreases to a half the width of that of the previous grid level ($h(l-1)$).
A new finer grid is generated whenever the minimum local Jeans length $\lambda_J$ 
decreases below $8\,h(l_\mathrm{max})$ (thus the so-called Jeans condition of \citealt{truelove1997}
is always satisfied in our simulation).
We set the maximum and minimum grid level to $l_\mathrm{max}=10$ and $l_\mathrm{min}=1$.
The grid of $l=10$ has a spatial coverage of 250\, AU and a cell width of 2\,AU.
Then our calculation can examine the first core and its vicinity with 2\,AU at the highest resolution.

We adopt a critical Bonnor-Ebert sphere of $M=0.6\,M_\sun$, $R=2000$\,AU with a uniform
kinetic temperature $T_\mathrm{kin}=10$ K as an initial condition.
A uniform magnetic field of $B=4.2\times 10^{-5}$ Gauss threads the initial 
critical Bonnor-Ebert sphere.
Initially the cloud rotates rigidly with dimensionless angular speed $\omega\equiv 
\Omega_0/(4\pi G\rho_0)^{1/2} = 0.3$ (where $\Omega_0 = 5.4\times 10^{-13}$
s$^{-1}$ is the initial angular speed and $\rho_0 = 3.8\times 10^{-18}$ g cm$^{-3}$
or $n_0 = 10^6$ cm$^{-3}$ is the initial central density).
We increased the initial density by 70\% to induce gravitational collapse (see for more 
details, \citealt{machida2008}).

For the thermal evolution, \citet{machida2008} employed the barotropic relations $P(\rho)$ 
based on the equation of state of a one-dimensional radiative hydrodynamic simulation study of \citet{masunaga2000}.
In this paper, we perform a non-LTE line transfer simulation with a numerical scheme 
optimized for a uniform Cartesian grid as a postprocessing procedure, 
making use of a snapshot data of the hydrodynamic 
simulation result of the nested grid scheme.
In order to maintain a balance between preserving information obtained by the 
nested grid simulation having a fine grid in the innermost region and grasping a large scale feature of 
a whole system,  we added a slight modification to the equation of state.
In particular, we artificially raised the polytropic index after $n>10^{14}$ cm$^{-3}$ to $\gamma=2$, and 
stopped further contraction beyond $n>10^{14}$ cm$^{-3}$ (see Eq.[5] of \citealt{machida2008}
for comparison).
The following set of barotropic relations $P(\rho)$ was used in the form of a piecewise power-law,
 \begin{equation} 
P = \left\{
\begin{array}{ll}
c_s^2 \rho & \rho < \rho_c, \\
c_s^2 \rho_c \left( \displaystyle\frac{\rho}{\rho_c}\right)^{7/5} &\rho_c < \rho < \rho_d, \\
c_s^2 \rho_c \left( \displaystyle\frac{\rho_d}{\rho_c}\right)^{7/5} 
  \left( \displaystyle\frac{\rho}{\rho_d} \right)^{2} & \rho_d < \rho, 
\label{eq:eos}
\end{array}
\right.  
\end{equation}
where $c_s = 190$ m s$^{-1}$ is sound speed at $T_\mathrm{kin} = 10$\,K, $\rho_c
=1.92\times 10^{-13}$ g cm$^{-3}$ ($n_c = 5\times 10^{10}$ cm$^{-3}$) and 
$\rho_d = 3.84\times 10^{-10}$ g cm$^{-3}$ ($n_d = 1.0\times 10^{14}$
cm$^{-3}$), respectively. 
Due to this modification, 
the evolution of low speed outflows from just outside the first core can be 
calculated for a relatively long timescale.
A similar technique was used in the numerical study of \citet{tomisaka2002}.

\begin{center}
\begin{figure}[htbp]
\epsscale{0.6}
\plotone{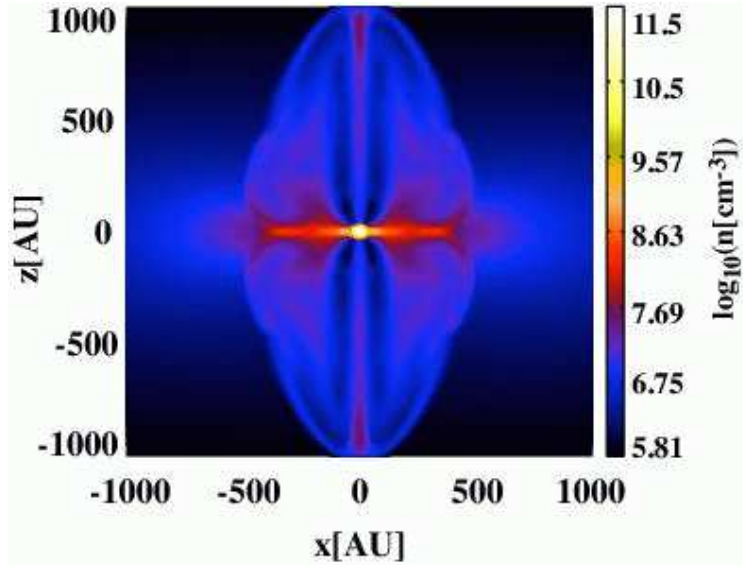}
\caption{Density structure of the snapshot data used in radiative transfer simulation.
The density at $x-z$ plane at $y=0$ is displayed.
Bipolar outflows extend along the $z$-axis ($R\lesssim 1000$\,AU), 
and the first core is seen as a quite compact dense 
region ($R\lesssim$ 100AU) at the center.
Additionally a geometrically-thick protostellar disk appears as an extended region 
with a scale height $\approx 200-300$ AU.
In the vicinity of the launching regions of the outflows, the density structure shows 
a cavity surrounding the central axis along the $z-$axis, and a shell-like 
dense regions of the shape of a letter ``U" surrounding them.}
\label{fig:densty}
\end{figure}
\end{center}

As is described in detail in the next subsection (\S\ref{sect:rt}), we calculated emission
from the model YPO separately from the magnetohydrodynamic simulation.
We made use of a part of ($l=7$) snapshot data of magnetohydrodynamic simulation 
at $t=4200$ yrs after the formation of the first core 
as an standard input to the radiative transfer code for a uniform Cartesian grid. 
Figure \ref{fig:densty} shows the density structure used in the radiative transfer simulation.
It reveals bipolar outflows extending to about 2000 AU, and a very compact and dense 
first core at the center.
In the vicinity of the launching region of the outflows,  a cavity structure exists around
the central axis ($z-$axis), and a shell of a shape of a letter ``U" surrounding it.
At this stage the major axis of the first core is about $\lesssim 50$ AU. 
In addition to the outflows, one can observe a disk-like structure with a radius $R\lesssim 1000$
AU and a scale height $\lesssim$ 200 - 300 AU.
This disk-like structure also appears in the calculations of \citet{machida2008} after 
a protostar forms. 
As described above, we did not explicitly calculate the evolution of the central region where
density exceeds 
$n\gtrsim 10^{14}$ cm$^{-3}$. 
However, our long-term calculation effectively enables an examination of 
further evolution of accretion phase to a protostar over a large scale (Figure\,\ref{fig:densty}), 
though lack of spatial resolution
prevents us from observing a protostar formation in a much smaller spatial scale than
the minimum cell size ($\sim$ 10 AU).
This disk-like structure is a protostellar disk in its earliest stage, 
and will evolve to a flatter and rotationally supported circumstellar disk 
(hereafter we refer to it as a ``protostellar disk"; \citealt{hennebelle2008}).
To study the formation and evolution processes of a rotation-supported circumstellar disk we 
obviously need calculations of an even longer term evolution with higher resolution
than the snapshot time adopted here.
Since this article focuses on the larger scale of young molecular outflows so that 
emission distribution can reasonably be resolved by the ALMA telescope, such an  
exceptionally high resolution simulation is beyond the scope of this article.
The protostellar disk is characterized by both infall and rotational motions.
In order to examine how the emitting and/or absorbing matters outside the outflows affect the emergent 
line intensities, it is convenient to define regions of different velocity 
structure, i.e., by the ratio of infall and rotation velocities.
Hereafter in this article we denote the least dense gas surrounding the outflows and the 
protostellar disk as an ``envelope",  which has only a small ratio of rotational to 
infall motion and low density ($n\lesssim 2.24\times 10^6$ cm$^{-3}$) in the 
computational domain.

\subsection{Radiative Transfer Simulations} \label{sect:rt}

We calculate continuum and molecular line emission individually using the snapshot data
of magnetohydrodynamic simulation described in the previous subsection (\S\ref{sect:hydro}).
For a continuum emission, we calculated thermal emission from dust grains with opacity
of \citet{hildebrand1983} \citep[see also][]{beckwith1990}.
For molecular lines, we performed three-dimensional non-local thermodynamic 
equilibrium (non-LTE) radiative transfer simulations with an algorithm based on \citet{hogerheijde2000}
\citep[see][]{wada2005}.
In the numerical calculation of radiative transfer, rate equations in statistical equilibrium, 
\begin{eqnarray}
  n_J\sum_{J^{\prime}\ne J} R_{JJ^{\prime}} &=& \sum_{J^{\prime}\ne
  J}(n_{J^{\prime}}R_{J^{\prime}J}),
  \label{eq:rateeq} \\
    R_{JJ^{\prime}} &=& \left\{
                \begin{array}{ll}
                A_{JJ^{\prime}}+B_{JJ^{\prime}}\bar{J}
		   +C_{JJ^{\prime}} &  J > J^{\prime},  \\
                B_{JJ^{\prime}}\bar{J} + C_{JJ^{\prime}} & J < J^{\prime},
                \end{array} 
                \right.  \label{eq:rateeq2}
\end{eqnarray}
are solved in each cell in the simulation box.
In equation (\ref{eq:rateeq2}), $A_{JJ^{\prime}}$ and $B_{JJ^{\prime}}$ are 
Einstein's coefficients for spontaneous decay and radiative transitions, 
and $C_{JJ^{\prime}}$ is for collisional transition rates from the energy level $J$ to
$J^{\prime}$, respectively.
Since almost all the hydrogen is in molecular form at this evolutionary stage, we can safely
take molecular hydrogen as a collisional partner, and then $C_{JJ^{\prime}}$
becomes
\begin{displaymath}
  C_{JJ^{\prime}} = n_{\mathrm{H}_2}\gamma_{JJ^{\prime}},
\end{displaymath}
where $n_{\mathrm{H}_2}$ is number density of molecular hydrogen, and 
collisional coefficient $\gamma_{JJ^{\prime}}\approx \langle v\sigma \rangle$ is 
taken from the database tables of Leiden University (LAMDA; \citealt{schoier2005}).
In order to calculate the mean intensity appearing in equation (\ref{eq:rateeq2}) $\bar{J} \equiv (4\pi)^{-1}\int I_\nu d\Omega$, we used a set of sampling rays passing through each 
cell with a random direction determined by Monte Carlo method.
Typically the number of sampling rays for each cell is 220 in the simulations.
As is inferred from Figure\,\ref{fig:densty}, radiation field becomes anisotropic, so we 
examined if the number of ray ($N_\mathrm{ray}=220$) is sufficient to incorporate 
the anisotropic radiation field by experiments with larger number of $N_\mathrm{ray}$ 
up to 420. 
We found only negligibly small differences in these results with 
relative precision less than 0.1\%, and the central region where coincidence becomes 
worse than 10\% for these runs 
occupies only a small fraction of $10^{-6}$ for low  $J$ lines and a few percent for the maximum $J$ ($=10$) in volume.
We hence concluded $N_\mathrm{ray}=220$ is sufficient to obtain reasonably correct 
results, and below we show results of $N_\mathrm{ray}=220$ runs.

The specific intensity along the sampling rays are calculated by integrating a standard
radiative transfer equation, 
\begin{equation}
\frac{dI_\nu}{d\tau} = -I_\nu + S_{\nu},  \label{eq:radtrans}
\end{equation}
where $S_\nu$ is the source function, and $\bar{J}$ in a cell is obtained 
from the average of specific intensities along the sampling rays passing through the cell.
The excitation temperature is calculated by iterative calculations of equations (\ref{eq:rateeq2}) 
and (\ref{eq:radtrans}) until a self-consistent solution of energy level population $n_J$ 
and mean intensity $\bar{J}$ is obtained.
We repeat the iterative calculations until the energy level populations $n_J$ converge
with relative precision of $\sim 10^{-6}$ between the present and former iterations 
\citep[see for details][]{yamada2007}.
As shown below (\S3), optical thickness becomes rather large in most of the lines we calculated. 
To avoid misunderstanding of slow convergence caused by the huge optical thickness
as the final convergence of self-consistent radiation field, we calculated 
until the relative difference of $(i-1)$-th and $i$-th energy level populations fell below 
$10^{-11}$ or less as an experiment (which is taken to be $10^{-6}$ as a canonical 
value in the present paper). 
Comparison 
with this longer calculation did not find any significant difference 
from the results with relative precision of $10^{-6}$, and thus we concluded 
that the level of convergence $10^{-6}$ is precise enough to avoid false convergence.

Kinetic temperature of the snapshot result of hydrodynamic simulation is almost  
isothermal with a low temperature $T_\mathrm{kin}$ = 10K, but in a small region 
at the center $T_\mathrm{kin}$ reaches as high as 177 K.
Then in order to include the possible energy level cascade from high-$J$ levels
\citep[see for details, Appendix of][]{yamada2007}, 
we examined two choices of maximum energy level $J_\mathrm{max}=10$ and 16. 
We found these two runs gave negligible difference of 0.1\% or less for 
the levels $0\le J \le 10$. 
Therefore in order to save computational time, we chose the maximum energy level 
$J_\mathrm{max} = 10$ in all the runs in this article.
Furthermore we confirm our solution by running two cases for each parameter set with 
different initial energy level populations of LTE and optically thin limit \citep{vandertak2005}.
We accept only solutions that meet the criteria that runs of two initial level population
coincide with relative precision of $\lesssim 10^{-4}$ for both of $n_J$ and 
source function $S_\nu$. 
The coincidence of $S_\nu$ is necessary because of the radiative transition terms 
in equation (\ref{eq:rateeq2}), or photon trapping effect (\citealp{yamada2009}).

We assume a Gaussian profile $\phi_\nu$ for absorption coefficient $\alpha_\nu$ 
only with pure thermal broadening 
$\Delta\nu_\mathrm{th}\equiv \nu_0/c\sqrt{(2k_BT_\mathrm{kin}/m)}$, 
\begin{mathletters}
\begin{eqnarray}
  \phi(\nu)d\nu &=& \frac{1}{\Delta\nu_\mathrm{th}\sqrt{\pi}}
     \exp{\left[ -\frac{(\nu - \nu_0)^2}{(\Delta\nu_\mathrm{th})^2} \right]} d\nu, \\
  \int_0^{+\infty}\phi(\nu)d\nu &=& 1, 
\end{eqnarray}
\end{mathletters}
where $\nu_0$ is the frequency of local line center, and $k_B$ is the Boltzmann constant, 
respectively.
Doppler effect due to bulk velocity of fluid elements is incorporated through 
the frequency shift of locally estimated $\nu_0(\boldsymbol{v})$ in each fluid element.
Observations of young YSOs such as Class 0 objects and molecular cores suggested 
relatively small turbulent motion compared with thermal width from the line profile
analysis \citep{difrancesco2001, belloche2002, bourke2005}.
It is also observationally suggested that turbulent motion is subsonic 
in isolated starless cores \citep{myers1983, goodman1998,caselli2002}.
The snapshot data we use in our radiative transfer simulation is at the 
evolutionary stage on the way from a quiescent starless core to a Class 0 object.
Then we neglect turbulent broadening in the absorption profile $\phi(\nu)$.

The snapshot data from the magnetohydrodynamic simulation cover only a part of 
the computational domain spatially as well as temporally.
The magnetohydrodynamic simulation described in \S\ref{sect:hydro} is calculated with a 
fine resolution ($dx = 2$ AU) by nested grids from $l_\mathrm{min}=1$ up to $l_\mathrm{max}=10$, 
but we used a set of $l=7$ grid data as a standard.
The spatial size of the numerical box of $l=7$ grid is $\approx$ (2000AU)$^{2}\times$ 1000AU, 
which covers
a whole outflow (see Figure \ref{fig:densty}). 
Precise calculation of emergent emissions from the YPO 
under consideration requires
absorption and/or emission of the region outside of the $l=7$ grid.
However in this article, we assume the Cosmic Microwave Background (CMB) 
radiation of $T_b = 2.73$ K
as a background radiation for simplicity, instead of approximate inclusion of outer part.
Most of the lines in our calculations are optically thick, and thus 
the effect of this artificial assumption about background radiation are limited 
in a very narrow region close to the outer boundary of $l=7$ grid.
In order to estimate the effects of the outer envelope to the radiation field of  
$l=7$ grid, we calculated the identical radiative transfer simulation for $l=5$ grid data 
which covers a larger volume but has 4 times coarser spatial resolution than $l=7$ grid.
We confirmed that the mean line profiles of $l=5$ calculations averaged over the field of view 
do not show qualitative differences
from $l=7$ calculations especially for mid- and high-$J$ lines of SiO molecule. 
Although a full nested grid radiative transfer calculation is necessary to a precise  
evaluation, we conclude our results of $l=7$ grid data do not suffer from the 
significant influence of the outer boundary for SiO lines.
For $^{12}$CO and its isotopologue lines they turned out to suffer from the outer boundary effect.
Thus we will show $l=5$ grid data results as well as the canonical $l=7$ grid data 
results for $^{12}$CO and its isotopologues.

We calculated rotational transitions of the representative molecules used in 
observations of molecular outflows ($^{12}$CO, $^{13}$CO, C$^{18}$O and SiO).
Among these molecules, we show in this article the results of $^{12}$CO (and 
its isotopologues) as the 
most popular outflow tracer, and SiO as a representative of high density tracer
molecule since the average density in our snapshot data is high ($\sim 10^6$
cm$^{-3}$).
The purpose of this article is to illustrate the characteristic features in the emission 
in relation to the dynamical evolution of YPOs and their velocity fields, 
rather than chemical evolution.
So we assume a spatially uniform molecular abundance distribution.
SiO molecules are in general considered to be formed by gas-phase reactions of Si atom
desorbed from dust grain mantles in a shocked gas with high velocity $\gtrsim$ 
25 km s$^{-1}$ \citep{caselli1997,schilke1997}, and thus they are regarded as a 
high-velocity shock tracer.
However, in the early stage before frozen-out onto dust grains,  
SiO molecule can be formed in gas-phase reactions more than 10\% of the 
initial Si abundance in a short timescale $10^4$ yrs 
before frozen-out dust onto grains \citep{langer1990}.
Additionally, observation showed that in some outflows the location of the brightest SiO emission 
does not coincide with the location of the strongest terminal shock (e.g. HH 211: \citealt{hirano2006}), 
and lower-velocity components than 25 km s$^{-1}$ have been detected 
\citep[e.g.,][]{codella1999, gibb2004}.
These facts implies that the formation mechanism of SiO molecule 
in protostellar objects remains still ambiguous \citep{arce2007}.
Then in order to prevent possible uncertainties in a specific chemical evolution model, we 
treat the abundance distribution as a free parameter in the numerical experiments.
Since CS would be less sensitive to shock, and it has similar critical densities and wavelengths
as SiO, the emission of CS lines resemble that of SiO in our calculations with uniform 
abundance.
We will show CS results based on the chemistry models in future.
HCO$^{+}$ will also be examined in the future article as a charged molecule.
We take molecular abundances relative to hydrogen molecule as $y_\mathrm{CO}
=3\times 10^{-4}$ and $y_\mathrm{SiO} = 2\times 10^{-8}$ as  fiducial values.

\section{Results}

In this section we present our simulation results analyzed in usual methods 
in radio observation.
Spectral energy distribution (SED) of continuum emission,  maps of synthesized 
molecular line intensities integrated along the line-of-sight velocity, position-velocity diagrams, 
channel maps, and line profile maps are presented.
While we mainly consider high angular resolution observations with the ALMA telescope, 
we focus on the qualitative features in the synthesized radiation field because of the 
limitation of our simulation set-ups (see discussion in \S2). 
That is, we do not include any response of realistic observation instruments by,
for example, beam convolution and so forth.
We also neglect atmospheric or system noises for the same reason.

\subsection{Continuum: Thermal Emission from Dust Grains} \label{sect:continuum}

Theoretically, continuum observation is easier to achieve a higher signal-to-noise
ratio in high angular resolution observation in comparison to lines.
We calculated the thermal emission from dust grains in hope of revealing 
smaller internal structures compared to line emission.

\begin{center}
\begin{figure}[htbp]
\epsscale{1.0}
\plotone{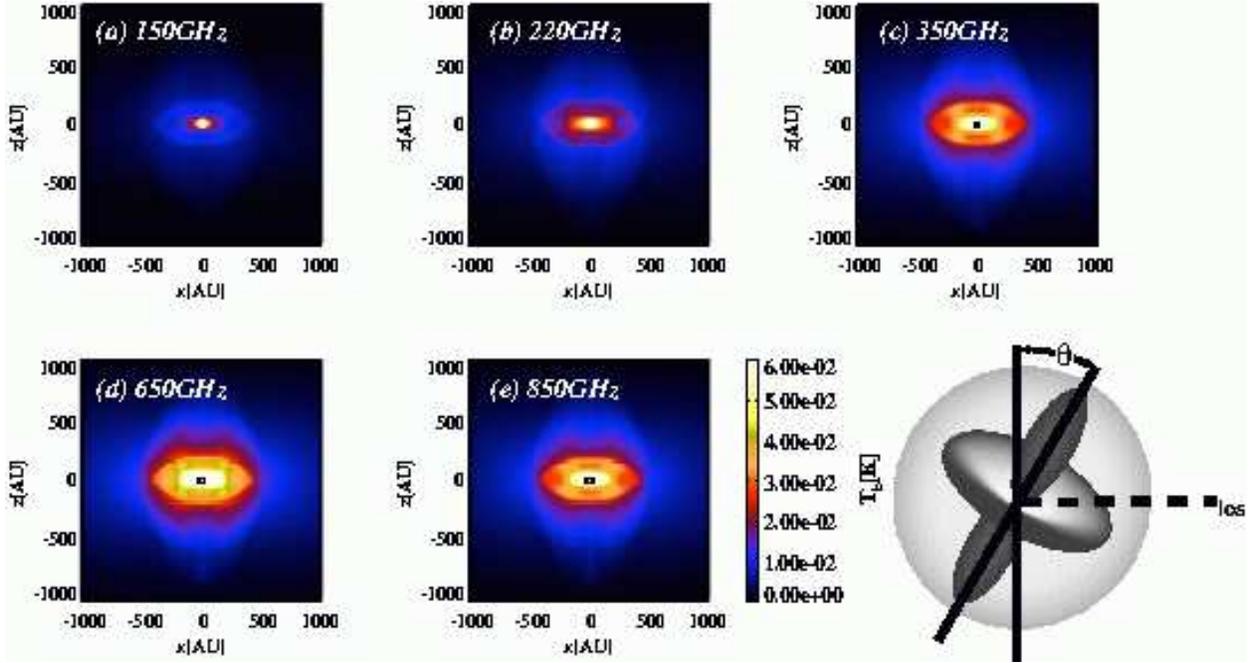}
\caption{Maps of thermal emission from $T_d=10$\,K dust grains. 
Five panels correspond to different wavelengths in the millimeter and submillimeter bands.
All these maps have an inclination angle $\theta=$30\degr\,(see the bottom right 
illustration for definition of the inclination angle $\theta$).
Black solid lines at the center ($R\sim 20-100$\,AU) of the panels (c), (d) and (e) 
indicate regions of which optical thickness exceeds 1.
Correction of finite optical thickness for conversion from $T_b$ to 
column density seems  unimportant except for an innermost region 
even in the highest frequency in the submillimeter band in our simulation, but
thick regions broaden their areas if the outer part of the simulation domain 
is included.}
\label{fig:column}
\end{figure}
\end{center}

In the snapshot under consideration, the density is as high as $n\gtrsim 10^6$ cm$^{-3}$, 
and we can assume a tight coupling between the gas and dust grains.
As the kinetic temperature of gas is nearly isothermal ($T_\mathrm{kin}=10$\,K) except for
a compact region around the first core at the center, we can take the dust 
temperature $T_d = T_\mathrm{kin} \simeq 10$\,K.
Then dust thermal emission is written as \citep{hildebrand1983,beckwith1990}
\begin{eqnarray}
  I_{\nu} &=& B_\nu\left\{1-\exp{(-\tau_\nu)}\right\}, \label{eq:cont}  \\ 
  \tau_\nu &=& \int \mu m_p n\kappa_\lambda dl,  \\
  \kappa_\lambda &=& 0.1\left( \frac{0.25 ~ \mathrm{[mm]}}{\lambda} \right)^{\beta} ~ ~
  ~ \mathrm{cm}^2 ~ \mathrm{g}^{-1},  \label{eq:kappa}
\end{eqnarray}
where $B_\nu$ is Planck function, $\mu$ is the mean molecular weight, $m_p$ is proton mass, $dl$ is 
a line element along the line of sight, respectively.
In our calculation we take $\beta=2$ as a fiducial value.
Figure \ref{fig:column} shows distributions of thermal emission 
of $T_d=10$\,K dust grains in terms of brightness temperature $T_b\equiv c^2/(2\nu^2k_B) I_\nu$.
Five panels correspond to different frequencies, $\nu=$ 150, 220, 350, 650, and 850
GHz falling in the band ranges of ALMA receivers.
We take a single value of inclination angle of the outflow axis to the plane of the sky 
$\theta=$30\degr\,(see a bottom right illustration for definition of $\theta$) in Figure \ref{fig:column}.
We found that average physical quantities such as mean intensity do not strongly depend 
on $\theta$, except for the morphology of intensity contours that follow the 
column density distribution.

The region bright in dust emission is mainly the central part of the protostellar disk
that has high column density up to $\lesssim 10^{23}$ cm$^{-2}$.
Panels (c), (d), (e) of Figure \ref{fig:column} show a weak structure of ``X"-shape extending
from the rim of a box-shape image of the protostellar disk.
These are base of the slow speed outflows of a wide opening-angle (see Figure\,\ref{fig:densty}).
These structures along the outflows are seen clearer in closer view to the edge-on 
of the protostellar disk (i.e. small $\theta$), and are hidden by the geometrically thick 
protostellar disk at $\theta\gtrsim$ 60\degr\,(close to the pole-on view).
Note that since the intensity range shown in Figure \ref{fig:column} is restricted for a clear view, 
the contrast between intensities of the outflow cavity wall and the protostellar disk is 
more emphasized compared with the true values.
Feasibility of detection of the signatures from the outflow cavity in 
dust emission depends on several factors actually, such as sensitivity of a telescope and 
gas-to-dust ratio in the outflow, and requires more detailed investigation.
Figure \ref{fig:column} shows that the brightness temperature of dust continuum 
emission is not strong ($\sim 0.02$ K for 350 GHz), but a huge bandwidth (two 
polarization channels of 8 GHz) of 
receivers of the ALMA telescope enables the detection of these emission easily: 
for 2 mK sensitivity, the required observation time amounts to about 2.6 hours for 
an object at the distance of 140 pc and -30\degr\,declination with 0.3 arcsecond resolution.
\footnote{We used the ALMA sensitivity calculator for the evaluation throughout the paper 
({\tt http://www.eso.org/sci/facilities/alma/observing/tools/etc/}). }
For 850 GHz with 1.0 mK sensitivity, required time will be about 8.8 hours, which is 
possible and reasonable to carry out.

Figure \ref{fig:column} shows that the emission of higher frequency is 
brighter, whereas equation (\ref{eq:kappa}) also implies an increase 
in optical thickness with frequency.
Finite optical thickness moderates the increase in intensity (Eq.\,[\ref{eq:cont}]) compared
to the optically thin regime.
In Figure \ref{fig:column} (c), (d) and (e), we indicate the thick ($\tau_\nu\ge 1$) regions 
of $R\sim 20-100$\,AU with solid black contours at the center.
One can see that though $\tau_\nu$ increases with frequency, the region of $\tau_\nu \ge 1$, in which correction is needed in conversion from $T_b$ to column density, 
is confined in a very small region at the center.
As described in \S\ref{sect:hydro}, we have to consider the outer region
surrounding the $l=7$ grid employed in our calculation for a more precise estimation.
Table \ref{tbl:maxtau} summarizes the maximum and the mean optical thicknesses 
in the field-of-view defined by the computational domain ($l=7$ grid).
It also shows values for  $\beta=$ 1.5 and 1.

\begin{center}
\begin{figure}[htbp]
 \epsscale{0.4}
 \plotone{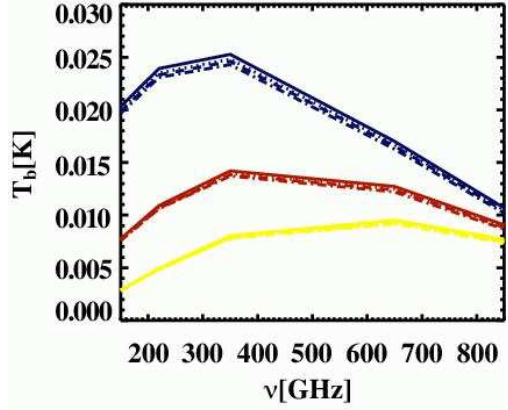}
 \caption{SED averaged over the field-of-view defined in the computation domain.
 Blue lines are for $\beta=2$, red lines are $\beta=1.5$, and yellow lines are 
 $\beta=1$, respectively.
 For a value of $\beta$, solid line indicates the viewing angle of $\theta=$90\degr
 (pole-on view),  dotted line is for $\theta=$60\degr, dashed line is for $\theta=$30\degr, and 
 a dot-dashed line is for $\theta=$0\degr (edge-on view).}
 \label{fig:cont_sed}
\end{figure}
\end{center}

Continuum emission shows symmetric distribution with respect to the $x=0$ axis 
($y-z$ plane) and $z=0$ axis ($x-y$ plane), reflecting symmetric density structure
(Figure\,\ref{fig:densty}).
This symmetry is greatly different from line intensity distribution discussed below 
(Figure\,\ref{fig:integ}).
Panels (c) and (d) of Figure \ref{fig:column} show very weak emission from the narrow
structure along the outflow axis.
Although the dust emission can have information of the density structures other than
the protostellar disk, it will be quite difficult to detect these weak emissions because of 
much brighter protostellar disk.
In spite of our expectation of high angular resolution continuum observations, 
actual observation of dust emission will work as a measure of inclination angle of 
a protostellar disk and configuration of outflows, rather than examining the 
tiny launching region of outflows at the center.
This is similar to what is done in current observations, indicating that  
we would not take advantage of the high resolution of the ALMA telescope. 
We plot average SEDs for three values of $\beta$
(2, 1.5, and 1) in Figure \ref{fig:cont_sed}.
Since the protostellar disk is geometrically thick, mean intensity averaged over the 
field-of-view is almost independent of the inclination angle $\theta$ as well as azimuthal
angle around the outflow axis $\phi$, thus multi-band observations will be
able to clearly determine $\beta$.
Differences in SED by $\beta$ appear stronger in longer wavelength regime where
optical thickness effect is minimal.

\subsection{Line Emission: A Probe of Velocity}  \label{sect:line}

In this subsection we show the results of non-LTE line transfer calculations.
In contrast to the continuum emission, lines become a key to derive 
information of the velocity field which describes the characteristic features 
in the youngest evolutionary stage in star formation.

\subsubsection{Excitation Status}

\begin{figure}[htbp]
\epsscale{0.6}
\plotone{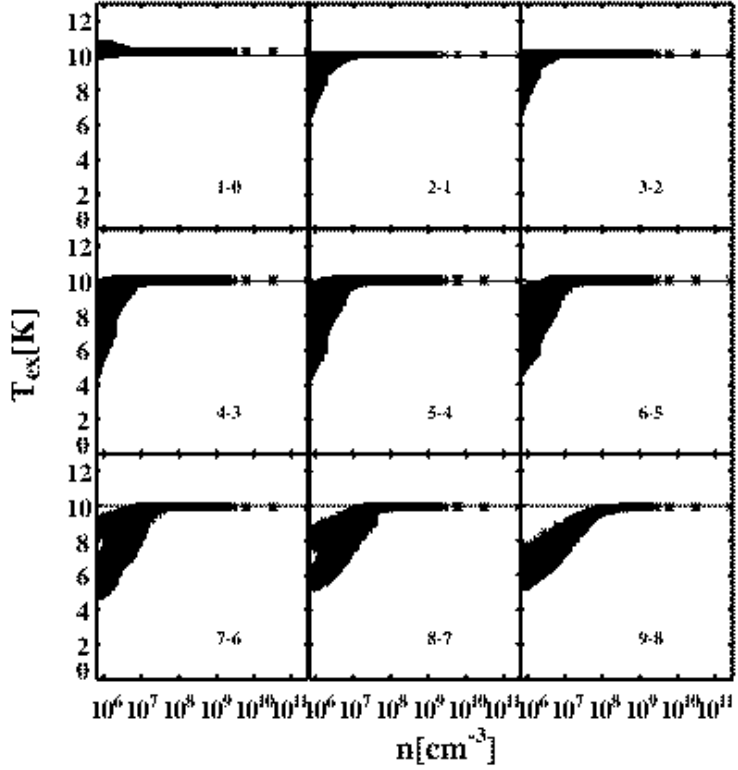}
\caption{Excitation temperature of SiO line in each cell is plotted as a function
of density of the cell. 
Thin horizontal lines at $T_\mathrm{ex}=10$\,K indicate a reference for 
thermalization in an isothermal gas of $T_\mathrm{kin}$ of 10K.
Low excitation transitions in millimeter bands ($J=1-0, 2-1$) are almost fully 
thermalized, but the excitation conditions of high excitation transitions of 
which critical density for LTE rapidly increases with $J$ ($n_\mathrm{crit}
\approx 10^5\times J^3$ cm$^{-3}$), depart
from LTE in the low density regime ($n\lesssim 10^7- 10^8$ cm$^{-3}$).}
\label{fig:Tex_SiO}
\end{figure}

In this subsection we examine excitation temperature.
In Figure \ref{fig:Tex_SiO} we plot the excitation temperatures 
($T_\mathrm{ex}(ij)\equiv -h\nu_{ij}/k_B\{\ln(g_i/g_j)-\ln(n_i/n_j)\}^{-1}$, where 
$g_i$ is statistical weight and $h\nu_{ij} \equiv E_i-E_j$) of SiO 
in each cell as a function of local density.
Figure \ref{fig:Tex_SiO} shows that low-$J$ transition in millimeter band the energy
level population is almost fully thermalized ($T_\mathrm{ex} \simeq T_\mathrm{kin}
\simeq 10$\,K), while high-$J$ transition (upper level of $J$, $J_\mathrm{upp}\gtrsim 4$) in 
submillimeter band departs from LTE in the low density regime ($n\lesssim 10^7 - 10^8$
cm$^{-3}$).
This is due to the increasing critical density for LTE with $J$ ($n_\mathrm{crit}\approx
10^5\times J^3$ cm$^{-3}$ for SiO rotational transitions).
The low density region ($n\lesssim 2\times 10^6$ cm$^{-3}$) roughly corresponds to 
the envelope and a part of the protostellar disk, and the high density region ($n\gtrsim
10^8$ cm$^{-3}$) concentrates to the first core and a part of the midplane of the protostellar 
disk (see Figure\,\ref{fig:densty}).
Therefore, for example, we can expect that $J=4-3$ ($\nu_{43} = 173.883100$\,GHz) 
lines will show non-LTE effects in the envelope, and $J=7-6$ ($\nu_{76}=
303.9268092$\,GHz) will show them in the envelope and the protostellar disk.
We do not show a similar figure of $^{12}$CO: because of lower critical density
of $^{12}$CO compared to SiO ($n_\mathrm{crit}\approx 10^2\times J^3$ cm$^{-3}$), 
energy level population of $^{12}$CO is almost completely thermalized up to the 
highest transition ($J=10-9$) in our line transfer calculations.

\subsubsection{Integrated Intensity}

\begin{center}
\begin{figure}[htbp]
\epsscale{0.8}
\plotone{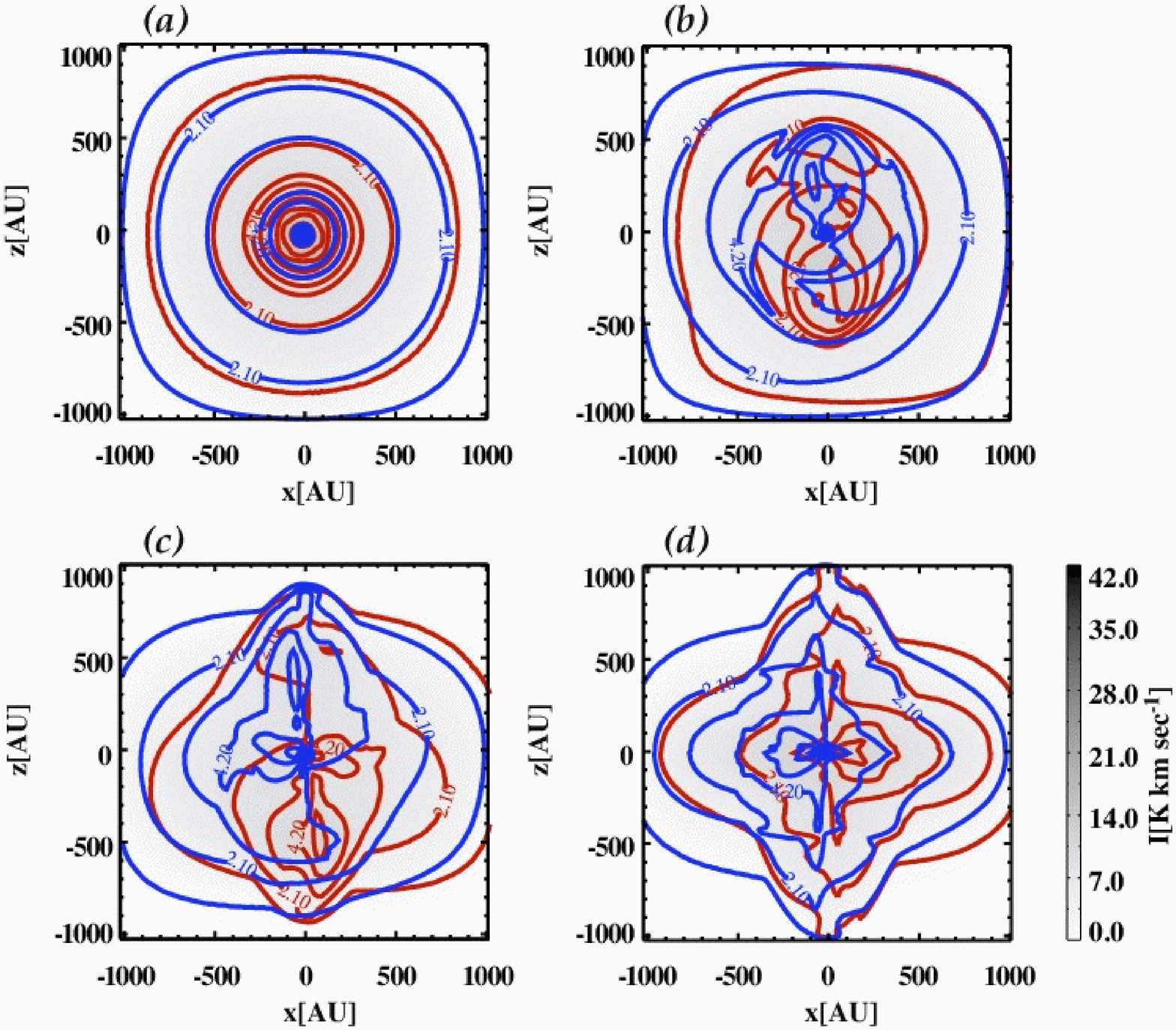}
\caption{Maps of integrated intensity of SiO$(7-6)$ line (gray scale). Four panels correspond to 
different viewing angles:  panel (a) is $\theta=$90\degr, (b) is $\theta=$60\degr, 
(c) is $\theta=$30\degr, and (d) is $\theta=$0\degr.
Red and blue contours indicates integrated intensities in $v_r>0$ (red) and $v_r<0$ (blue)
components, respectively, with equal contour spacing of 1.05 K km s$^{-1}$.}
\label{fig:integ}
\end{figure}
\end{center}

For a brief overview of the synthesized line emission distribution, first we show 
maps of integrated intensity of SiO$(7-6)$ in Figure \ref{fig:integ}.
In Figure \ref{fig:integ}, we overplot red and blue components in color lines
on the total integrated intensity in gray scale, where intensities of total, 
``red component", and ``blue component" are defined with the line-of-sight projection
of the relative velocity with respect to the rest of the object $v_r$:
\begin{eqnarray}
  I_\mathrm{tot} &=& \int^{+\infty}_{-\infty} T_b dv_r, \\
  I_\mathrm{red} &=& \int^{+\infty}_0 T_b dv_r, \\
  I_\mathrm{blue} &=& \int^0_{-\infty} T_bdv_r,
\end{eqnarray}
respectively.
Four panels in Figure \ref{fig:integ} are for different inclination angles ($\theta=$90\degr, 
60\degr, 30\degr, and 0\degr\, for panels (a), (b), (c) and (d), respectively).
Distributions of red and blue components can be described by superposition of 
rotation of the protostellar disk and the outflows, infall and outflow motions.
The rotational motion causes symmetric distribution of red and blue components 
with respect to the $x=0$ axis
(except for $\theta=$90\degr\, or pole-on view: see panels (c) and (d) of Figure \ref{fig:integ}), 
and infall and outflow motions give rise a symmetric distribution of them around the 
center (i.e., $I_\mathrm{red}$ is almost the same with $I_\mathrm{blue}$ at the location 
rotated by 180\degr\, around the center: panels (b) and (c) of Figure \ref{fig:integ}).
Outermost contours surrounding the projected outflows in panels (b) and (c) become 
circular, because the envelope   
has nearly radial infall velocity.

\begin{center}
\begin{figure}[htbp]
\plotone{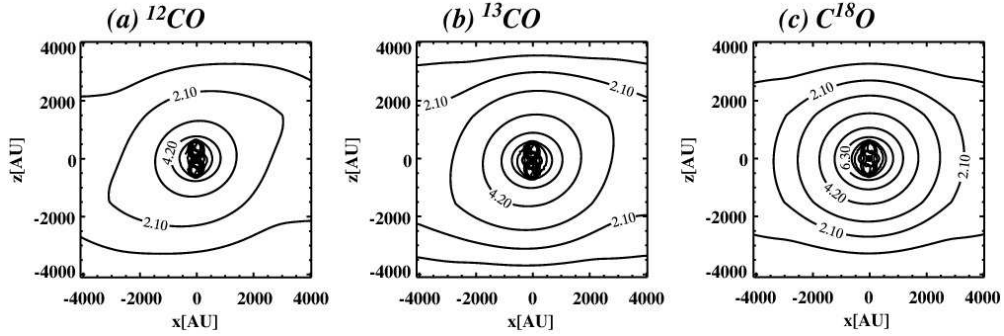}
\caption{Maps of integrated intensity of $^{12}$CO, $^{13}$CO, 
and C$^{18}$O $(1-0)$ lines taken from the $l=5$ data calculations. 
Because of the use of $l=5$ grid data, the area of the field-of-view is 16 times larger than that  
in Figure \ref{fig:integ}. Contours are drawn with equal spacing of 1.05 K km s$^{-1}$.}
\label{fig:integ_co}
\end{figure}
\end{center}

Next we show the integrated intensity of $^{12}$CO, $^{13}$CO, and C$^{18}$O in Figure
\ref{fig:integ_co}. 
As is discussed in \S\ref{sect:rt}, these molecules are more easily excited in the envelope 
compared with SiO, and hence the outer envelope surrounding the $l=7$ grid cannot 
be neglected. 
Line transfer simulations of lower $l$ grid data show that the photospheres 
of lines of these molecules are located just inside the radius of the initial Bonnor-Ebert sphere.
We show $l=5$ grid data results in Figure \ref{fig:integ_co} ($\theta=$60\degr), 
which cover the whole initial Bonnor-Ebert 
sphere and thus the photospheres of CO and its isotopologues. 
In these figures we adopted molecular abundances $y$ to be $2\times 10^{-4}$, 
$2\times 10^{-6}$, and $2\times 10^{-7}$ for $^{12}$CO, $^{13}$CO, and C$^{18}$O, 
respectively. 
Figure \ref{fig:integ_co} shows that because of the large optical thickness (see Figure\,\ref{fig:avepro_co} 
below), intensity distributions have only weak structure compared with SiO (Figure\,\ref{fig:integ}).
This result means that, due to the high density of the parent molecular core, even the 
rare isotopologue lines of CO becomes optically thick as well as $^{12}$CO. 
In more evolved system, difference in the optical thicknesses in $^{12}$CO and 
its isotopologues can be used to trace 
different components, but they are 
not good probe of emission signatures of the very young embedded YPO like that modelled in 
this article.
Because the timescale of CO freeze-out is short in the high density core ($\sim 10^4$ yrs
for $\gtrsim 10^6$ cm$^{-3}$: e.g., \citealt{flower2005}), it may be possible that significant 
amount of gas phase CO is locked on the dust grains in the youngest YPO.  
If extremely effective depletion reduces the molecular abundance in the
gas phase by a factor of $10^{-3}$ or less in the entire core, 
the signature of the embedded YPO would be 
easier to appear, but such a large factor of depletion is unlikely 
\citep[see chemistry model, e.g.,][]{aikawa2001}.
If the YPO further evolves, emission from the central protostar will heat up the dust grains and 
desorb CO into gas phase again, and furthermore the dissociative background radiation 
might prevent CO molecule from freeze-out \citep[e.g.,][]{jorgensen2005, jorgensen2006}. 
Taking into account these factors, the extremely strong CO depletion is unlikely in 
a real YPO, and our conclusion that CO and isotopologue lines are not useful to 
examine the kinematics of a YPO does not change, though absolute intensity and 
locations of photospheres would be slightly affected. 

\subsubsection{Average Line Profiles}

\begin{figure}[htbp]
\plotone{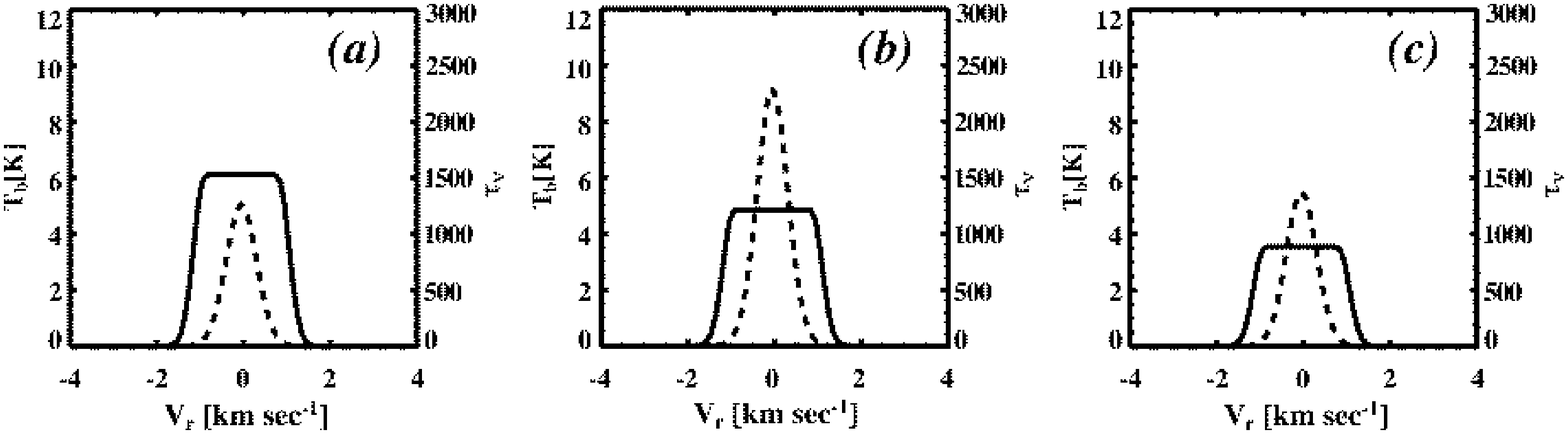}
\caption{Line profiles of $^{12}$CO averaged over the field-of-view (solid lines) 
and average $\tau_\nu$ profile (dashed lines) as a function of line-of-sight velocity for 
$\theta=$60\degr\,view for $l=5$ grid.
Panels (a), (b), and (c) are for different transitions, $J=1-0$, $2-1$, and $3-2$, respectively.
Because of the huge optical thickness, line profiles take a strongly saturated profile
with very weak wing signatures ($|v_r|\lesssim 2$ km s$^{-1}$).}
\label{fig:avepro_co}
\end{figure}

Figure \ref{fig:avepro_co} plots line profiles of $^{12}$CO emission averaged
over the field-of-view for $l=5$ grid calculations.
Three panels show different transitions ($J=1-0$, $2-1$, and $3-2$) for a fixed 
viewing angle $\theta=$60\degr.
All three transitions display a strongly saturated line profile with very weak 
wing emission ($|v_r|\gtrsim 2$ km s$^{-1}$).
The strongly saturated profile is due to two facts: 1) in our computational domain 
the density is higher than critical densities for these lines ($n_\mathrm{crit}
\sim 10^2\times J^3$ cm$^{-3}$) and the energy level population is 
fully thermalized up to the maximum $J$ (=10) in our calculation,
and 2) resultant optical thickness becomes huge ($\tau_\nu \sim 1500 - 2500$
at the rest frequency; note the optical thickness becomes even larger if high resolution 
larger $l$ data are correctly incorporated) as plotted with dashed lines in Figure \ref{fig:avepro_co}.
Considering possible depletion of $^{12}$CO molecules on dust grains in 
the YPO, we performed numerical experiments with lower abundances
($y_\mathrm{CO} =  3\times 10^{-5}$ and $3\times 10^{-6}$) using the same 
snapshot data. 
However, we could not find significant differences in line profiles even with 
a 100 times smaller molecular abundance.
These imply that in our calculation of a YPO, energy 
level population of $^{12}$CO is governed by collisional transitions and 
by a relatively small contribution from radiative transitions ($\propto B_{JJ^{\prime}}$) in spite of large 
optical thickness.
The same results apply to isotope molecules like $^{13}$CO and C$^{18}$O.
Note that as the very young protostellar object in our calculation evolves
and forms a less dense and larger molecular outflow, optical thickness of 
$^{12}$CO will significantly be reduced and the saturated profiles will disappear.
The saturated profile in our results, which is not seen in current observations, 
appears because we calculated the emission from the compact high density
region at the center of the computational domain.

\begin{figure}[htbp]
\plotone{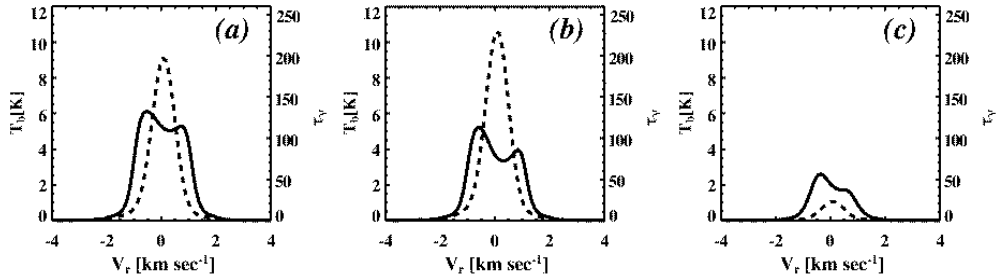}
\caption{The same figures as Figure \ref{fig:avepro_co}, but for SiO.
Panels (a), (b), and (c) denote $J=2-1$, $4-3$, and $7-6$ transitions, respectively.
Mean optical thickness reduces by a factor of $\sim 100$ compared 
to $^{12}$CO, and line profiles
take a double-horn shape skewed to the blue side.}
\label{fig:avepro_sio}
\end{figure}

In Figure \ref{fig:avepro_sio} we plot average profiles of SiO lines in a similar way
but for the canonical $l=7$ data results.
Since the optical thickness diminishes by a factor of $100$ or more compared to
$^{12}$CO, the 
average profiles do not saturate but appear as a double-horn shape.
Obviously SiO lines are better tracers to probe kinematics in the YPO  
under consideration compared with $^{12}$CO (and its isotopologue) lines, 
besides chemical evolution in the very early stage that has to be further investigated 
to conclude whether SiO is the {\em best} tracer. 
The double-horn profiles appear in all the maps with different inclination angles $\theta$,
including pole-on view with various degree of skewness in red and blue and 
varying depth of the dip at $v_r=0$.
From these arguments and Figures \ref{fig:avepro_co} and \ref{fig:avepro_sio},
radiative transitions ($\propto B_{JJ^{\prime}}$) comparatively contribute in
determining the energy level population of SiO molecule as collisional
transitions.
Like in the case of $^{12}$CO, we performed numerical experiments of 
radiative transfer calculations with several lower molecular abundances  
down to $y_\mathrm{SiO} = 2\times 10^{-10}$, and confirmed that 
both intensities and profiles of SiO lines change according to the molecular abundance.
Our results demonstrate that interpretation of subthermally excited lines 
along with chemical evolution should be done carefully.

\subsubsection{Intensity Weighted Mean Velocity along the Lines of Sight} \label{sect:vel}

\begin{center}
  \begin{figure}
  \plotone{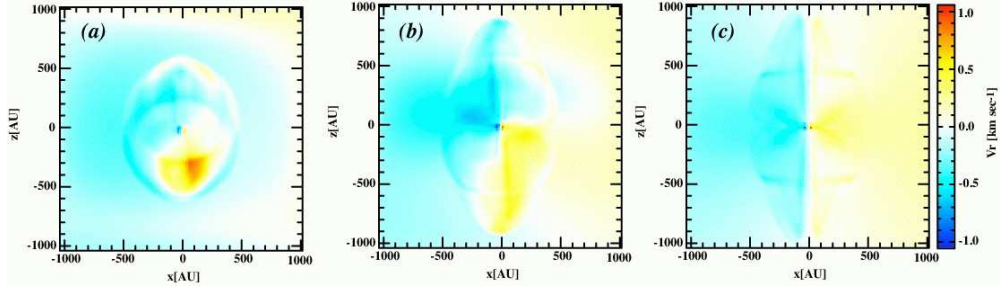}
  \caption{The velocity first moment ($\langle v_r\rangle$) 
  maps of SiO$(7-6)$ line.
  Panels (a), (b), and (c) are different inclination angles ($\theta=$60\degr,
  30\degr, and 0\degr, respectively).}
  \label{fig:sio76vel}
  \end{figure}
\end{center}

To display velocity structure clearer, we calculated the intensity-weighted
mean velocity along the line of sight (or first order moment),
\begin{equation}
 \langle v_r \rangle \equiv \frac{\displaystyle\int T_b v_r dv_r}{\displaystyle\int T_b dv_r},
\end{equation}
and show the distribution of $\langle v_r\rangle$ of SiO$(7-6)$
line in Figure \ref{fig:sio76vel}.
As is seen in Figure \ref{fig:avepro_sio}, in our calculations the optical thickness of SiO
is also large, though smaller than $^{12}$CO.
Then the intensity-weighted mean velocity $\langle v_r\rangle$ roughly traces 
velocity at $\tau \simeq 1$ region near the outer boundary, and hence it does not 
depend strongly on $J$. 
In Figure \ref{fig:sio76vel} three panels with different viewing angles ($\theta=$60\degr,
30\degr, and 0\degr) are presented.
As is seen in Figure \ref{fig:sio76vel}, $\langle v_r\rangle$ appears antisymmetric 
around the center in panels of $\theta=$60\degr\,and 30\degr\,cases (i.e., 
$\langle v_r\rangle \simeq -\langle v_r\rangle$ at the position rotated by 180\degr\, around
the center).
There appears a quite characteristic pattern of peaks of 
$\langle v_r\rangle$ in ``S" shape in these panels.
These features originate in the rotation of the outflows around the 
outflow axis, which are driven by magneto-centrifugal force in the 
vicinity of the first core \citep[e.g.,][]{tomisaka1998,tomisaka2000,machida2008}.
If the outflow having a cavity rotates and inclines with respect to 
the plane of the sky, the total velocity component along the line of sight 
consists of flowing velocity along the outflow axis and rotation 
around the outflow axis.
Opposite sign of rotation velocity in two sides of the outflow axis 
generates the gradients of $\langle v_r\rangle$ across the (projected) 
outflow axis with opposite signs in red and blue lobes.
If the outflow axis is exactly in the plane of the sky, thus generated gradients 
in $\langle v_r\rangle$ maps appear with the same sign in the two lobes
directing upwardly and downwardly, 
because the protostellar disk and the outflows rotate in the same direction.
One can see absence of the ``S"-shape in the $\theta=$0\degr\,map 
(panel (c) of Figure\,\ref{fig:sio76vel}), where the outflow axis aligns exactly in the 
plane of the sky.

\begin{center}
  \begin{figure}
  \plotone{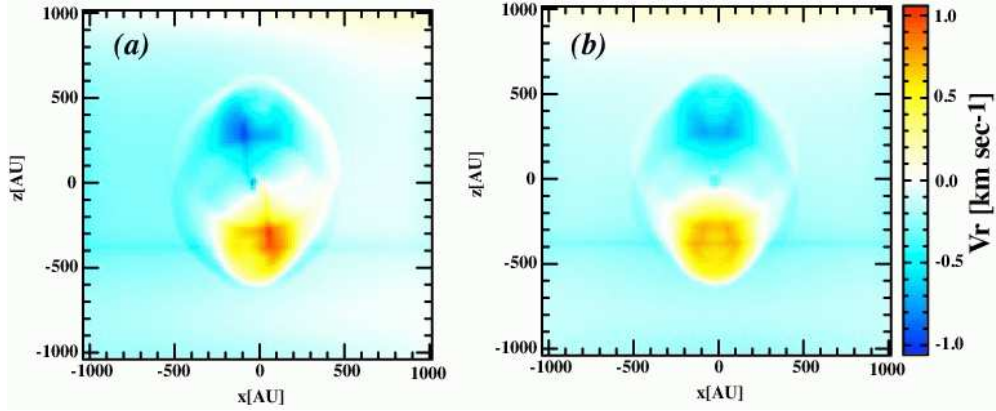}
  \caption{The velocity first moment $\langle v_r\rangle$ maps of SiO$(4-3)$
  line for $\theta=$30\degr\,view. 
  Panel (a) is taken from the original result with $v_\phi\ne 0$, and panel (b)
  is a result of $v_\phi=0$ experiments.
  Symmetric and anti-symmetric distributions of peaks are due to rotation of 
  the outflows and the protostellar disk.
  Similar trend is observed in all the lines we calculated.}
  \label{fig:sio43vel}
  \end{figure}
\end{center}

In order to demonstrate the effects of rotation of the outflow on 
line emissions clearer, we performed 
line transfer simulation with a modified snapshot data. 
We artificially set $v_\phi(\equiv\sqrt{v_x^2+v_y^2}) = 0$ in the snapshot and 
put it as an input to the line transfer calculation.
Figure \ref{fig:sio43vel} (b) shows the results of $v_\phi=0$ run of SiO$(4-3)$ line.
The velocity gradients across the outflow axis disappear in the $v_\phi=0$ run
(panel (b) in Figure\,\ref{fig:sio43vel}), while the original result of SiO$(4-3)$
($v_\phi\ne 0$) conserves the similar gradients as in Figure \ref{fig:sio76vel}. 
These velocity gradients will be able to observationally confirmed with 
a spatial resolution of $\approx 100-200$ AU (required resolution depends 
on the inclination angle) and a velocity resolution of $\approx 0.1$ km s$^{-1}$.
If a YPO of this evolutionary stage is at a distance of 140\,pc, 
100 AU corresponds to 0.71$^{\prime\prime}$, and above structure could be 
resolved with current instruments such as the Submillimeter 
Array.

\begin{center}
  \begin{figure}
  \plotone{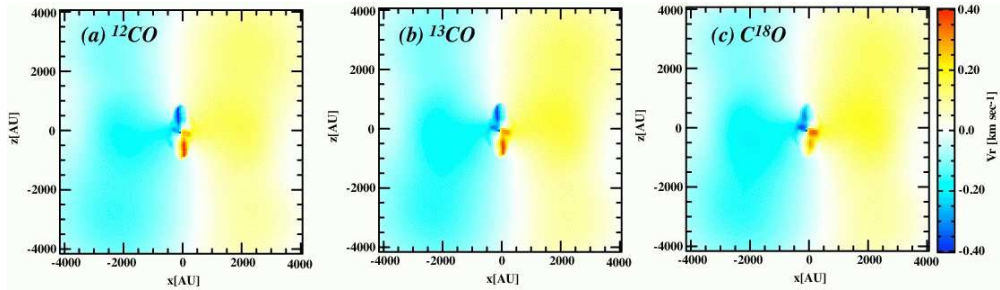}
  \caption{The velocity first moment ($\langle v_r\rangle$) 
  maps of $^{12}$CO$(3-2)$ and its isotopologue line for $l=5$ calculations.
  Panels (a), (b), and (c) are $^{12}$CO$(3-2)$, $^{13}$CO$(3-2)$, and C$^{18}$O, 
  respectively, and have the same viewing angle $\theta=$60\degr. The global 
  symmetric gradient of $\langle v_r\rangle$ would originate in the rotation of the 
  parent core.}
  \label{fig:co21vel}
  \end{figure}
\end{center}

In the case of CO, because of the locations of photospheres distant from the 
embedded outflow, $\langle v_r\rangle$ appears more weakly than SiO.
Figure \ref{fig:co21vel} shows the similar $\langle v_r\rangle$ maps with those 
in Figure \ref{fig:sio76vel}, but for $l=5$ data calculations. 
Three panels in Figure \ref{fig:co21vel} show $J=2-1$ transition of $^{12}$CO
($y=2\times 10^{-4}$), $^{13}$CO ($y=2\times 10^{-6}$), and C$^{18}$O 
($y=2\times 10^{-7}$) for a viewing angle $\theta=$30\degr, respectively.
One can observe that these maps of $^{12}$CO and its isotopologue lines quite 
resemble each other, and they present the weak velocity gradient at the central (2000 AU)$^2$ 
region, of which pattern is qualitatively similar to those in Figure \ref{fig:sio76vel}.
Smaller range of $\langle v_r\rangle$ compared to SiO is either because of greater optical depth 
or smoother density structure in the lower resolution $l=5$ grid data compared 
with the $l=7$ data. 
This figure again shows that CO and isotope molecules cannot distinguish different components 
based on the difference in optical thickness because of the high density 
in the model, and puts more emphasis on the disability of CO molecules to 
probe embedded YPOs.

\subsubsection{Position-Velocity Diagrams} \label{sect:pv}

\begin{figure}[htbp]
\epsscale{0.7}
\plotone{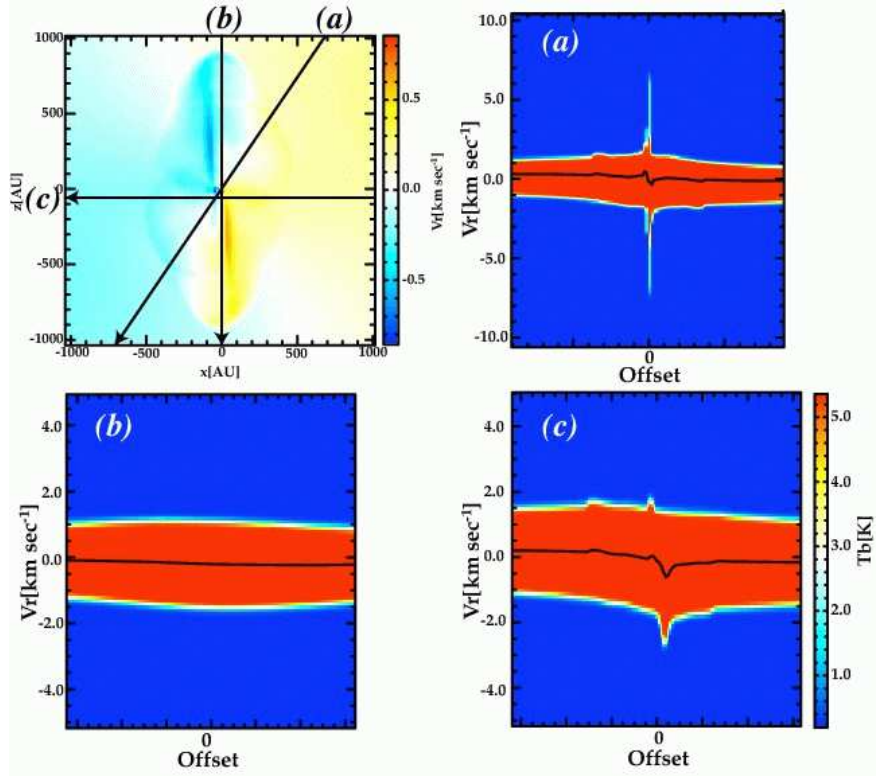}
\caption{Position-velocity diagrams of $^{12}$CO$(2-1)$ line along three cuts 
indicated on the top-left panel ($\langle v_r\rangle$) for $\theta=$30\degr\, view
(data are taken from $l=7$ grid calculation for a clear view).
Width of position-velocity diagrams are scaled to the length of the cuts from the 
center.
Solid lines indicate $\langle v_r \rangle$ along each cut.}
\label{fig:co_pv}
\end{figure}

We made position-velocity diagrams from the same synthesized data cubes.
First we show position-velocity diagrams for $^{12}$CO$(2-1)$ and $\theta=$30\degr\,in Figure \ref{fig:co_pv} for $l=7$ data for a clear view (we confirmed that $l=5$ results are not
significantly different from those of $l=7$ calculations).
All three panels in Figure \ref{fig:co_pv} show almost no structure (we will discuss
the origin of a spiky structure in the panel (a) below).
These featureless structures in position-velocity slices can be easily inferred from
the saturated shape of averaged intensity in Figure \ref{fig:avepro_co}.
In other words, velocity structures of the hydrodynamic data vanish in the 
structureless $T_\mathrm{ex}$ distribution and huge optical thickness even 
in the non-zero velocity part ($|v_r|\lesssim 2$ km s$^{-1}$).
Panel (c) of Figure \ref{fig:co_pv}, which is a cut passing through the 
region about 50 AU offset from the center, shows a similar but weaker spiky structure
as in panel (a). 
The origin of these spikes in panels (a) and (c) is the same (see below).
The cut of panel (b) passes through the center, and because of the large optical
thickness, it has only a weak velocity gradient.

\begin{center}
\begin{figure}[htbp]
\epsscale{1.0}
\plotone{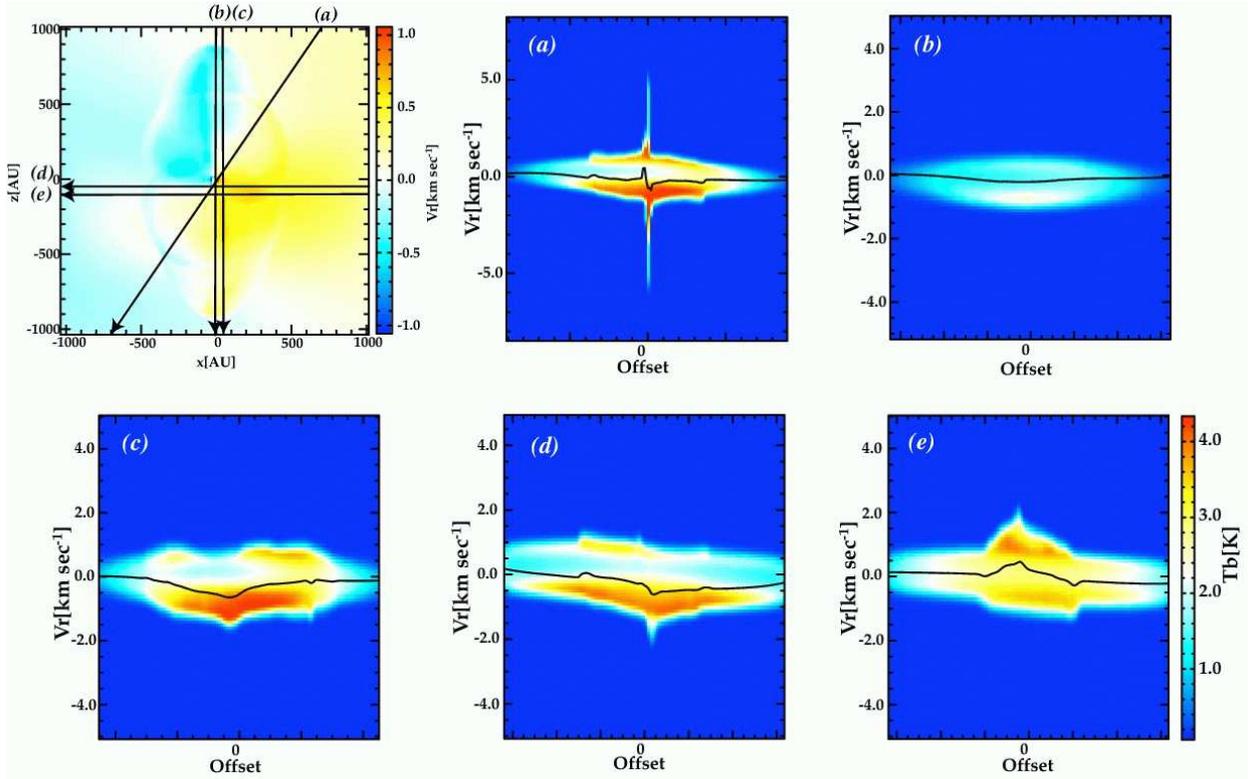}  
\caption{Position-velocity diagrams of SiO$(7-6)$ line. Horizontal width of 
diagrams are similarly scaled with the cut length on the $\langle v_r\rangle$
map at the top-left.
Compared with Figure \ref{fig:co_pv}, SiO line shows a variety of structures in
these position-velocity diagrams due to the non-LTE population.}
\label{fig:sio_pv}
\end{figure}
\end{center}

On the other hand, SiO lines which have much smaller optical thicknesses 
present a variety of structures in position-velocity diagrams due to the  
non-LTE energy level population (Figure\,\ref{fig:sio_pv}, SiO$(7-6)$ and $\theta=$30\degr).
The most characteristic features are: 1) some of these diagrams show a 
``gap" around $v_r =0$ km s$^{-1}$ (panels (a), (c) and (d)),  
2) a pair of cuts at the close position ($\lesssim 50$\,AU) along the same direction, such as 
a pair of (b) and (c),  presents drastically
different structures in corresponding position-velocity diagrams.
A pair of (d) and (e) is another good example.
These features are combination of a variety of densities (or emissivities) of the 
(relatively dense) outflows and the protostellar disk that have different bulk velocities.
Position-velocity diagrams can be used in combination with velocity 
channel maps (\S\ref{sect:channel}) and line profiles (\S\ref{sect:lprofile}) to interpret velocity field.
These variety of structures in position-velocity diagrams are obviously useful 
to reconstruct the original velocity and density fields from emission lines, although requires high
angular resolution observation.

Finally we mention the origin of the spikes of $\Delta v_r\sim \pm 5$ km s$^{-1}$ in
panel (a) in Figures \ref{fig:co_pv} and \ref{fig:sio_pv}.
These arise from the high velocities in the vicinity of the first core at the center and 
thermal broadening.
Kinetic temperature in the adopted snapshot is almost isothermal of low 
$T_\mathrm{kin} =10$\,K, but in the limited region close to and in the first core 
$T_\mathrm{kin}$ rises rapidly as high as 177\,K at a maximum, which 
corresponds to the thermal width $\Delta v_\mathrm{th} \sim 2$ km s$^{-1}$.
In our numerical model, the 
magneto-centrifugal force associated with the ``hour-glass" shaped magnetic fields formed 
around the rotating first core drives the outflow of a wide opening angle \citep[e.g.,][]{tomisaka1998, 
tomisaka2000, machida2008}.
In this model, magnetic field is straight on the $z-$axis (the outflow axis) 
and thus the magneto-centrifugal force does not work there.
The velocity near the $z-$axis is not outward, but inward.
At the center where gravitational potential depth is at its maximum, the inflow 
motion takes its maximum velocity as high as 1.8 km s$^{-1}$.
The combination of a wide thermal width and the highest velocity (in the whole 
snapshot data) results in spikes in panel (a) in Figures \ref{fig:co_pv} and 
\ref{fig:sio_pv} with a very large velocity range in the position-velocity
diagrams.
Because of the large thermal broadening,  velocity range at these spikes
becomes larger than the maximum bulk hydrodynamic velocity of the input data.
To confirm the origin of these spikes, we performed line transfer experiments 
with the same snapshot but its kinetic temperature is set to be 10\,K isothermal.
The spikes disappear in these experiments, as expected.

\subsubsection{Velocity Channel Maps} \label{sect:channel}

As an alternative expression of the velocity field imprinted in line emission, we 
show the velocity channel map of SiO$(7-6)$ line in Figure \ref{fig:channel1}.
In Figure \ref{fig:channel1}, extended diffuse components appearing in the
velocity range -0.6 km s$^{-1} \lesssim v_r \lesssim$ 0.6 km s$^{-1}$ are 
emission from the rotating protostellar disk.
The accreting envelope extending further outside does not contribute 
in the channel map, since its density ($n\simeq 10^6$ cm$^{-3}$) is smaller
than critical density of SiO$(7-6)$ ($n_\mathrm{crit}\simeq 1.3\times 10^7$
cm$^{-3}$ at $T_\mathrm{kin} = 10$\,K).
The density in the snapshot has a cavity-like structure and a slightly denser 
fan-shape structure surrounding it (see Figure\,\ref{fig:densty}).
However in the channel map (Figure\,\ref{fig:channel1}) the fan-shape structure 
close to  the outflow launching region 
is hardly observed because of the geometrically thick protostellar disk, even 
in the edge-on view.
Only in a very limited range of velocity channels and viewing angles can we 
see the fan-shape structure. 
This fan-shape structure is due to the magneto-centrifugal 
force driven flow from the rotating first core, and does not come from 
entrainment of the core matter by a high-velocity jet.
Our results mean that close attention should be paid in inferring a driving 
mechanism of outflows from observed channel maps especially when the 
target protostellar object is at the 
young evolutionary stage.

One of advantages of velocity channel maps is that it can demonstrate 
the rotational motions of the outflows more clearly compared to other diagrams.
For a comparison we plot a channel map expected for the identical dataset 
except for $v_\phi=0$ in Figure \ref{fig:channel2}
(see \S\ref{sect:vel}).
Comparison with Figure \ref{fig:channel1} clearly shows the existence or absence of 
rotation of the outflows.
High angular resolution observations with the ALMA telescope will reveal these 
complex structures of rotating outflows coupled with a protostellar disk.
Note that further imaging simulations are necessary to examine whether 
the diffuse emission from the protostellar disk can be correctly imaged with 
ALMA or other interferometers \citep*[e.g.,][]{kurono2009}. 
Tentative experiments showed that the total power array of the ALMA telescope 
is crucial to reproduce the diffuse emission in Figures \ref{fig:channel1} and \ref{fig:channel2} 
(Kurono \& Yamada, private communication).

\subsubsection{Line Profile Distributions} \label{sect:lprofile}

Finally we present an example of line profile distribution map in Figure \ref{fig:lprofile}
(SiO $(7-6)$ of $\theta=30$\degr\,view).
Figure \ref{fig:lprofile} shows that subthermal lines have double-horn or multi-peak 
profiles almost over the entire field-of-view.
A nearly flat profile at the map center comes from a combination of large bulk
velocity near the first core and a broad thermal width in the region (\S\ref{sect:pv}).
Blue component is relatively brighter on average (Figure\,\ref{fig:avepro_sio}), but  
opposite trend (brighter red component than blue one) appears locally.
This average blue-ward skewness would largely come from the overall contraction of the core
\citep[][]{zhou1995,evans1999}.

A protostellar object at the very young evolutionary stage has a quite complicated velocity 
structure composed of accretion of the parent molecular core, rotation, and 
outflow motion and so forth.
Therefore, as often referred in studies of spherical starless or prestellar cores
\citep{pav2008, tsamis2008}, the relation between emission features regarded 
as signatures of infall motions (e.g., blue-red asymmetry in line profiles) 
and actual infall motions is not necessarily well-established.
For example, in our calculations, the snapshot data has an infall motion in 
almost entire computational domain and optical thickness large enough for 
self-absorption,  nevertheless local enhancement of red intensity over blue  
is observed in some lines of sight.
\citet{adelson1988} found that rotation and expansion motions can 
show a similar double-horn profile with a brighter red component compared to a blue one. 
Interestingly, our simulation results show double-horn profiles even in the 
outer part, though weak,  where the optical thickness is not very large.
The asymmetric double-horn profile is basically a result of combination 
 of several factors such as: 1) blue-enhancement by asymmetric 
self-absorption due to the increasing 
$T_\mathrm{ex}$ toward the center in infall motion \citep{evans1999},  and 
2) bipolar velocity (infall, outflow, and rotation) reflected in the optically
thin regime.
An additional numerical experiment is performed in order for a simple examination 
of emissions from different velocity components.
We constructed two modified input data that have 1) radial infall velocity field  
outside the outflow regions ($\boldsymbol{v} = \boldsymbol{v}_\mathrm{r}=
\min(\boldsymbol{v\cdot e}_\mathrm{r}, 0)
\boldsymbol{e}_\mathrm{r}$, where $\boldsymbol{e}_\mathrm{r}$ 
is a unit vector in the radial direction; case A), 
and 2) the outflow region velocity only ($\boldsymbol{v}=(\boldsymbol{v}_\mathrm{out}
\cdot \boldsymbol{e}_z)\boldsymbol{e}_z$; case B).
In case B, the outflow region velocity $\boldsymbol{v}_\mathrm{out}$ 
is defined as the velocity having $|v_z|> 0.5$ km s$^{-1}$: this criterion turned out to be 
able to effectively capture the outflow region after several experiments.
Figure \ref{fig:lprofile_vr} shows the results of line transfer 
simulation using these artificially constructed data. 
Panel (a) of Figure \ref{fig:lprofile_vr} shows the result of case A. 
It shows blue-red infall asymmetry almost over the entire field-of-view and confirms 
this asymmetry in Figures \ref{fig:avepro_sio} and \ref{fig:lprofile} originates in overall 
infall motion of the core.
Panel (b) of Figure \ref{fig:lprofile_vr}, on the other hand, represents 
almost symmetric double-horn profiles especially in the vicinity of the projected outflow axis. 
Since in this result the infall velocity is artificially fixed to be 0 except for 
that on the $z-$axis (see \S\ref{sect:pv}), this symmetric profile would come from 
the bimodal outflow motion itself, and 
the skewed profiles close to the $z=0$ plane would be due to the protostellar disk.
In the full data set simulation, these components couple nonlinearly with 
structured emissivity, optical thickness,
and excitation temperature in different velocity and position. 
Further detailed analysis is necessary to disentangle them in a qualitative way.

\subsection{Velocity Structure and Line Emission}

As displayed in the previous subsection \S\ref{sect:line}, line emissions from YPOs are quite complicated.
In Figure \ref{fig:vel} we plot relations between bulk velocity and local density
 of the adopted snapshot data in the two-dimensional probability distribution function
 (PDF).
Figure \ref{fig:vel} shows that: 1) distribution of $v_\phi$ and $|\boldsymbol{v}|$ 
with respect to density is qualitatively similar, and 2) either of $v_\phi=0$ 
or $|\boldsymbol{v}|=0$ components are not significant in mass.
The velocity of the magneto-centrifugal force driven outflows 
are roughly equal to the Kepler speed \citep[e.g.,][]{kudoh1997}, and then 
in the adopted snapshot,  $v_\mathrm{rot}$, $v_\mathrm{inf}$ and 
$v_\mathrm{out}$ (rotation, infall and outflow speeds) become comparable.   
Hence the first feature in Figure \ref{fig:vel}, similarity in the 
distributions of $v_\phi$ and $|\boldsymbol{v}|$, 
represents comparable speed of $v_\mathrm{rot}$, $v_\mathrm{inf}$ and $v_\mathrm{out}$.
Taking into account the symmetric structure of the input snapshot 
with respect to the $z=0$ plane and the geometrically thick configuration of the protostellar disk,  
it explains a complex velocity structure 
that multiple velocity components with similar norms  
but different signs easily appear in the field-of-view (or on a line-of-sight) whichever 
inclination angle aligns the object. 
In a more evolved object such as seen in current observations, the velocity structure
becomes simpler mainly because a protostellar disk will evolve to a thinner rotationally
supported circumstellar disk with relatively small spatial extent compared to 
the outflows.
Hence a complex velocity feature is a manifestation of dynamical evolution of the star-forming 
object in the young evolutionary stage, when the outflows are embedded in the accreting 
envelope and the thick protostellar disk.
Therefore such complexity in emission induced by 
complex velocity structure is a characteristic of the early stage of star formation.

We plot the relation between bulk velocity of the snapshot data and excitation
temperature obtained in line transfer calculations in Figure \ref{fig:v_Tex} for 
SiO$(2-1)$ and $(7-6)$ lines as two-dimensional PDFs.
Figure \ref{fig:v_Tex} shows that $T_\mathrm{ex}$ has a wide scatter 
as a function of velocity, and hence self-absorption will be possible at various 
velocities.
Particularly,  a relatively large scatter of $T_\mathrm{ex}$ around $v_\phi=0$
and $v_z =0$ suggests a large possibility of self-absorption 
at zero velocity as well.
Either a small amount of matter with $\boldsymbol{v}=0$ (Figure\,\ref{fig:vel}) and
high probability of self-absorption at $v_\phi=0$ or $v_z=0$ (Figure\,\ref{fig:v_Tex}) can 
weaken the line emission at $v_r=0$ compared to the other velocity components.
This causes the gap at $v_r=0$ in position-velocity diagrams in Figure \ref{fig:sio_pv}.
The gap structure in position-velocity diagram were already seen in 
observation \citep[e.g.][]{hogerheijde1998, williams2003, lee2006}.
On the other hand, energy level population of $^{12}$CO is almost fully thermalized
in the nearly isothermal snapshot, thus $T_\mathrm{ex} \simeq T_\mathrm{kin}\simeq$ 10K 
and $^{12}$CO lines do not show a gap in 
position-velocity diagram (Figure\,\ref{fig:co_pv}).
Since neither Figures \ref{fig:vel} nor \ref{fig:v_Tex} include spatial correlation, 
they cannot explain all the features of emitted lines.
All the arguments above are statistical.
Further detailed analysis of spatial structures will appear in the following article.

\section{Discussion}

The early phase of protostellar objects and their evolution in its beginning 
are quite interesting and important problems in the study of star formation.
For the formation of low-mass stars, a model of \citet{adams1987} of spectrally 
categorized young stellar objects from Class 1 to Class 3 has been widely accepted.
While multi-dimensional studies using numerical methods have constructed realistic 
evolutionary models of molecular cores to protostars \citep[see e.g.,][]{tomisaka1998,
tomisaka2000,tomisaka2002, banerjee2006,machida2006,machida2008,hennebelle2008}, observational studies have 
proceeded on the basis of classical spherical models.
One of the difficulties in observational studies of the early stage of star formation is 
its deeply embedded signature in the parent molecular core. 
Observations have found a numbers of non-radial 
velocity fields in mostly spherical molecular cores  
\citep[e.g.,][]{belloche2002}, but they have been interpreted in terms of a 
simple superposition of arbitrarily introduced rotation or outflow/inflow motions within 
the framework of classical spherical models.
High angular resolution observation facilities like the ALMA telescope are expected 
to provide information of velocity fields even in small structures within a protostellar object.
However, as shown in the previous sections, the signals of emission from a YPO   
deeply embedded in the envelope are difficult to correctly decipher even with the high 
angular resolution of the ALMA telescope, because of the complex velocity field and 
resultant complicated optical thickness structure.
Furthermore identification of outflow itself would be 
far more difficult compared to the case of more evolved outflows currently observed (Figure\,\ref{fig:channel1}).
Therefore, careful investigation of how the true 
structures of objects are reflected in emission from them is necessary.

Recently the {\em Spitzer Space Telescope} found numbers of very low luminosity
objects (VeLLOs) in the mid-infrared band.
The distribution of molecular gas accompanying these objects and the existence or 
absence of outflow are quite uncertain (\citealt{bourke2006}; Yamada \& Ohashi, private communication).
If these VeLLOs are indeed related to the very early phase of star formation, 
irregular morphology of molecular emission might also be a natural 
consequence (e.g., Figure\,\ref{fig:channel1}). 
If future high-angular resolution observation of molecular emission of VeLLOs 
indicates an outflow signature,  
comparison of our results would provide a valuable insight 
on the earliest phase of star formation.
Among VeLLOs, for example, a molecular core L1521F in Taurus has a high 
central density, infall asymmetry, and is chemically evolved.
Since the discovery of a weak infrared source by the Spitzer telescope,
it is considered to be an object on the evolutionary way from an evolved starless 
core and Class 0 protostar stage \citep{bourke2006}, thus it would 
be a good 
candidate to be studied in the future high-angular resolution observation with
the ALMA telescope to test the prediction of our paper.
A more evolved VeLLO, IRAM 04191+1522 also in Taurus will be another good target of the 
ALMA telescope in the study of the early stage of star formation.
Besides VeLLOs, it might be possible that rotation signatures like 
Figure \ref{fig:channel1} were considered to be a manifest of other kinds of 
velocities in some of the observation of irregular outflows in young YSOs 
\citep[see e.g.,][]{difrancesco2001}.

\subsection{Relevant Radiative Transfer Simulation Studies}

Recently line transfer simulation studies of star formation have increased in number.
Many works on the early phase of star formation focus on chemical evolution 
in their one-dimensional modelling rather than on dynamical evolution.
\citet{tsamis2008} employed the inside-out collapse solution of \citet{shu1977} and calculated 
chemical evolution along with line transfer in collapsing cores.
They constructed a virtual 45-m class telescope and performed pseudo-observation of 
their simulation results.
\citet{pav2008} systematically examined the effects of rotation and infall motions 
on emergent line emission with their one-dimensional modelling.
Complex velocity fields (infall, rotation and outflow) in the early protostellar objects 
affect the emission line profiles in a quite non-trivial way (\S\ref{sect:lprofile}).
Generally blue-skewed asymmetry and inverse P-Cygni profiles are regarded to 
evidence an infall motion as well-understood mechanisms.
However, quantitative understandings of the effects of gas motion on line emission 
is not significantly well developed to solve an inverse problem.
In other words, we currently still cannot reconstruct 
the velocity field from the obtained line profiles to a satisfactory degree 
for a comparison with theoretical models, even for a spherical gas.
These works are experimental rather than applicable for asymmetric objects 
in real space, nevertheless they will form a powerful set of building blocks in 
interpreting more realistic simulation results like ours and observational results.

Many numerical works have been done in order to derive physical quantities
of outflows in observation data so that models can fit the observational results.
\citet{meyers-rice1991} performed numerical experiments of a simple concave model outflow, 
and \citet{rawlings2004} succeeded in constructing a best fit structure to the 
results of \citet{hogerheijde1998},  that consists of outflow 
cones and a surrounding envelope with their two-dimensional radiative transfer calculations.
These studies employed minimal models to reproduce the 
temporal structure imprinted in observations. 
On the other hand, in more realistic calculations presented in this article,   
the most difficult factor in the identification of the emission components from
the outflows is the contamination from the geometrically and optically thick protostellar disk (Figure\,\ref{fig:channel1}).
In a more evolved system, outflows cover wider spatial extents than the 
protostellar disk, and the protostellar disk will evolve to a flatter circumstellar disk, 
so that the outflows are observable without shielding of a thick disk.
In other words, the youngest outflows would not appear similar to those seen in current observation.
Interferometric observational studies proposed a 
disk/outflow model based on multi-components appearing in their data, but with an 
assumed geometrically thin disk.
Possible effects on the emission by a thick protostellar disk on the evolution toward  
a circumstellar disk is first mentioned in this article.
A thick protostellar disk appears in a number of numerical studies \citep{tomisaka1998,tomisaka2000,tomisaka2002,banerjee2006,machida2006,machida2008}.
Detailed study of emission is necessary for a verification of these results in future observations.

\subsection{Launch of Outflows}

Our calculations have a relation to outflow launching mechanisms.
One of the characteristic features of outflows driven by magneto-centrifugal force
is its rotation around the flow axis.
In contrast, if the outflow is accelerated by the momentum transfer from
the jet (so-called ``entrainment" mechanism), the molecular outflow would rotate very slowly.
This is expected by the fact that, in the entrainment mechanism, the
angular momentum is not effectively redistributed from the jet which has
a small angular momentum to the molecular outflow which has a large
moment of inertia.
Our line transfer calculations have shown that the rotation of the outflows appears in 
velocity channel maps, position-velocity diagrams and maps of $\langle v_r\rangle$.
\citet{launhardt2009} recently found a gradient of $\langle v_r\rangle$ supposed 
to present rotation of an outflow form a transition object from Class 1 to Class 2 (or younger), 
CB\,26 in their IRAM PdBI observation.
In addition to an evolved system like CB\,26, if rotation of outflows in younger 
objects will be discovered, they will be a good constraint for the models of 
outflow launching mechanisms.
The younger the object, the tighter the constraint will be, though a younger system
is embedded in an envelope of parent molecular core matter, and its definite
detection is more difficult.
We found a signature from rotation of the outflows in velocity channel map, 
position-velocity diagrams and distribution of $\langle v_r \rangle$.
Further analysis of the line profile distribution as well as those in Figure \ref{fig:lprofile_vr} 
will provide clearer indications of the rotation signatures, and will 
appear in a future article.

In the evolution model we employed in this article \citep{tomisaka1998,tomisaka2000,
tomisaka2002,machida2008}, 
outflows are driven just outside a first core.
The strongest support for their model will then come from the emission from the 
first core.
However, we cannot derive any strong conclusion about this, since the 
numerical modeling adopted in this article cannot resolve a geometrically 
thin shocked region at the surface of the first core, and cannot follow the 
shock chemistry inside such a thin region.
\citet{omukai2007} and \citet{saigo2009} calculated emission from a 
thin cooling shock surrounding a first core.
They consistently examined emission from the radiative shock and 
absorption/re-emission in the surrounding envelope by sacrificing the detailed 
dynamical evolution.
\citet{omukai2007} found that reprocessings in the thick envelope is 
significant in a spherical cloud, whereas \citet{saigo2009} showed that rotation flattens 
the cloud and reduces the column density along the rotation axis by more than one
order of magnitude.
Our three-dimensional calculations of longer wavelength emission, mainly for the 
ALMA telescope, showed that line emission can be seen through the 
wider velocity range due to the non-spherical motions and geometry.
Different from \citet{omukai2007} and \citet{saigo2009}, our calculations do not 
explicitly include an outer absorbing envelope (\S\ref{sect:rt}).
Though we conclude that the outer envelope would not qualitatively alter our results of
SiO lines of $l=7$ grid 
by a simple survey of line profiles in $l=5$ calculation,  
these effects in outer envelope absorption and optical thickness should be 
included in an advanced numerical scheme in the future.

\subsection{Comments on the Real Observations}

Finally we briefly comment on the real observations especially with 
the ALMA telescope. 
Our calculations are the appropriate means for comparing the results of MHD calculations 
with observations since we calculate the radiation explicitly.
This article focuses on the study of characteristic features of radiation field 
of a YPO, and does not include any limitations 
from observational instruments such as finite beam size \citep[compare with e.g.,][]{tsamis2008}.
Although our results are ideal, we can make a simple estimate of the necessary observation time by the 
ALMA telescope using the synthesized emission presented in this article.
If we assume a YPO at the distance of 140 pc and declination $\delta=-30$\degr, typical 
size of the characteristic features in the synthesized maps (Figures \ref{fig:integ}, \ref{fig:integ_co}, \ref{fig:channel1} and \ref{fig:channel2}) is $\gtrsim$ 10 AU, or $\sim 0.1^{\prime\prime}$
in angular size. 
From Figure \ref{fig:avepro_sio}, the brightness temperature of SiO$(7-6)$ line is $\sim$ 3 K, 
and line width is $\sim 2$ km s$^{-1}$ . 
Then if we invoke the velocity resolution to be 0.1 km s$^{-1}$, which is used in our 
calculation,  and 0.3 K sensitivity, the necessary exposure time amounts to $\sim 5100$ seconds 
or $\sim$ 14 hours. 
For the continuum, the required exposure time reduces to $\sim$ 2.6 hours for the 
same object but 2 mK sensitivity because of 
a large bandwidth of the receivers installed on the ALMA telescope (\S\ref{sect:continuum}). 
It is noted that one should be careful whether diffuse emission from the protostellar disk in 
Figures \ref{fig:channel1} and \ref{fig:channel2} can be correctly reproduced
in interferometer observations or not. 
The tentative results of such imaging simulation for the ALMA telescope suggest that in order to 
reproduce the diffuse component, the total power array is crucially important, and 
the above estimates assume the use of the full array (12 m, 7 m, and the total power array; Kurono 
\& Yamada, private communication).
Lower $J$ lines, of which critical densities are lower than the average density in
the protostellar disk and the envelope, appear even extended than Figure \ref{fig:channel1}.
Besides the limitation that comes from radiative transfer scheme (\S\ref{sect:rt}), 
imaging simulation for interferometers will be necessary for direct comparison
of our results with actual observations \citep{takakuwa2008, kurono2009}.

\section{Summary}

In this article we surveyed the characteristic features of the radiation field from 
a dynamically evolving object at an early stage of star formation, 
on the way from a prestellar (starless) core to a Class 0 object.
We have presented the most realistic calculation results obtained to date from a combination of
three-dimensional MHD simulation and non-LTE line transfer calculations.
We calculated the dust thermal emission and rotational lines of $^{12}$CO (and 
its isotopic molecules) and SiO 
molecules in the snapshot data consisting of the compact outflows ($R\sim 1000$\,AU), 
the protostellar disk and the envelope.
We found that:

1) Dust thermal emission traces a dense mid-plane in the protostellar disk rather than
very small structures in the launching region of the outflows.
A very weak fan-shape signature along the base of the outflows appears in viewing 
angles close to edge-on, but it is likely to be overwhelmed by the much brighter protostellar disk.

2) Because of the high density ($n\gtrsim 10^6$ cm$^{-3}$) in our snapshot data, 
$^{12}$CO and its isotopologue lines are fully thermalized and are not suitable to probe velocity field.
Our results are in apparent disagreement with current observation of more evolved outflows, 
but can be reasonably understood in terms of the different densities at different evolutionary
stages.
This implies that the energy level population of CO is governed by collisional 
transitions, and the molecular abundance does not significantly affect the results.
We experimented with lower molecular abundances to mimic possible depletion
of $^{12}$CO molecules,  and found no significant differences in the final results.

3) On the other hand, SiO lines are not completely thermalized because of their higher critical
densities ($n_\mathrm{crit}\sim 10^5\times J^3$ cm$^{-3}$), and are more suitable for  
probing the velocity structure in contrast to CO.
Line profiles of SiO are characterized by a double-horn shape over almost the entire field-of-view, and
position-velocity diagrams showed a variety of structures strongly dependent on
locations of different cuts.
These structures arise from a complex distribution of optical thickness and non-LTE
energy level populations induced by several velocity components, such as 
outflow, inflow, and rotation.
In a dynamically evolving system, these velocity components have similar speeds but
different directions, thus any line-of-sight can pick up multiple velocity components of 
similar norms but different signs irrespective of viewing angles.
In order for a correct interpretation of such complex velocities from line emission, 
further detailed analysis is necessary.

\begin{center}
\begin{figure}[htbp]
\epsscale{0.8}
\plotone{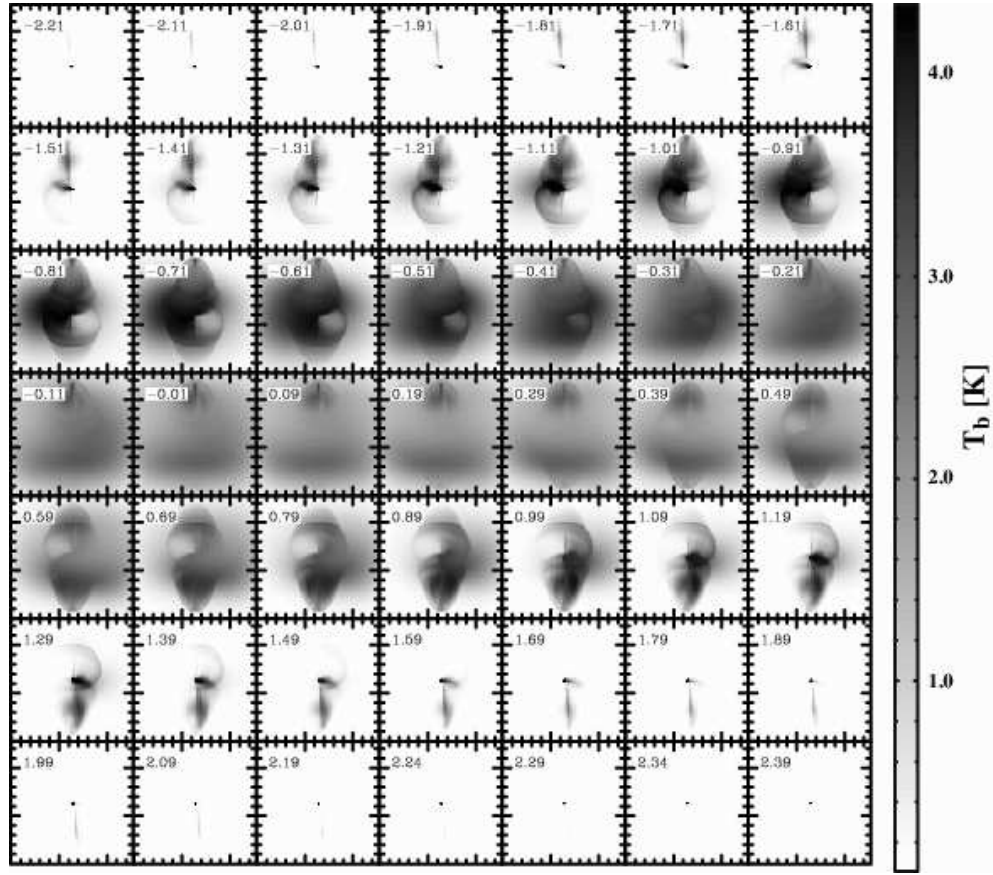}
\caption{Velocity channel map of SiO$(7-6)$ line for $\theta=$30\degr\,view.
The relative velocity in unit of km s$^{-1}$ from the rest of the object 
is labeled at the top-left of each map.
Diffuse components appearing in $|v_r|\lesssim 0.6$ km s$^{-1}$ are emission from
the protostellar disk.
Rotation of the outflows and the protostellar disk is clearly observed.}
\label{fig:channel1}
\end{figure}
\end{center}

\begin{center}
\begin{figure}[htbp]
\epsscale{0.8}
\plotone{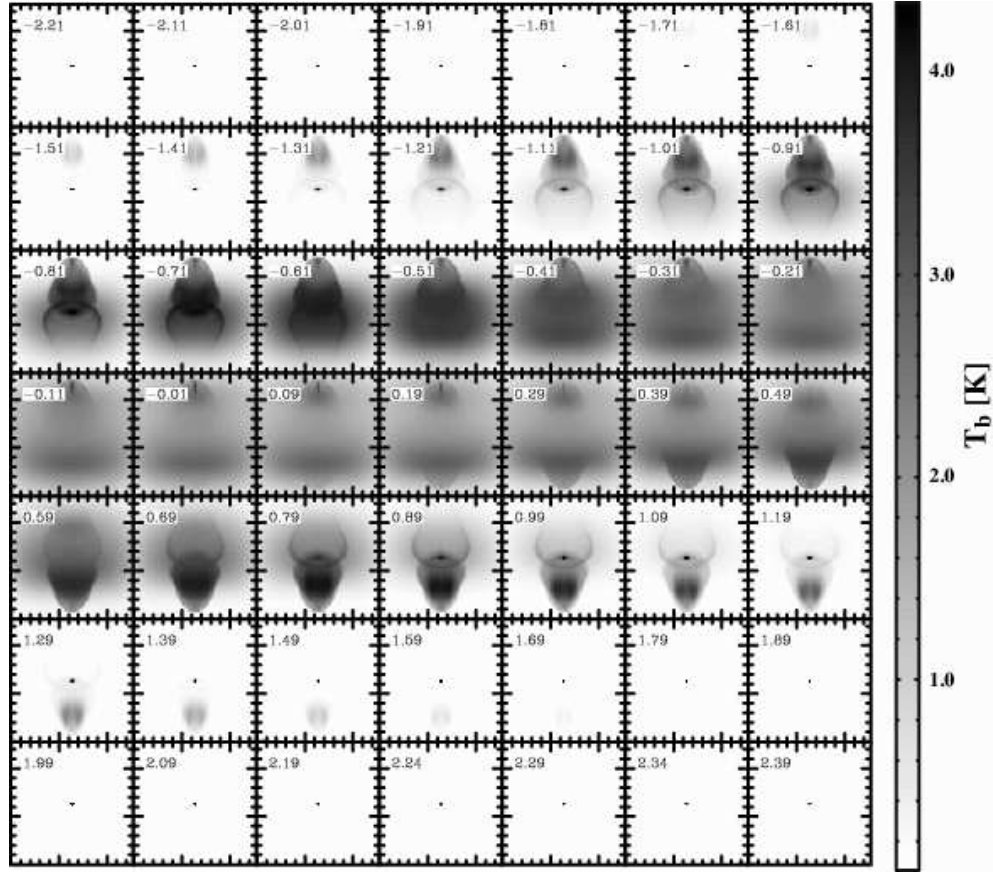}
\caption{Same as Figure \ref{fig:channel1}, but for a $v_\phi=0$ run.
No rotation signature from the outflows and the protostellar disk is observed.}
\label{fig:channel2}
\end{figure}
\end{center}

\begin{center}
\begin{figure}[htbp]
\epsscale{0.8}
\plotone{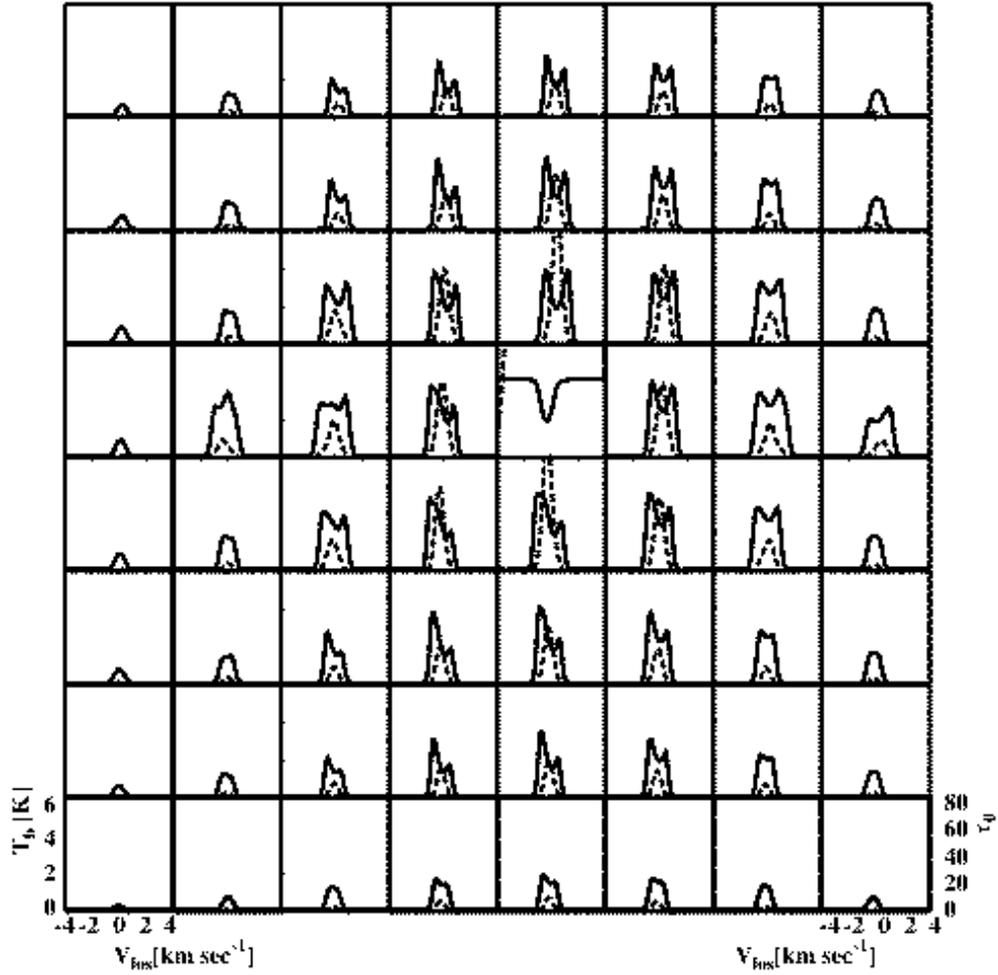}
\caption{Line profile distribution of SiO($7-6$) for a $\theta=$30\degr\,view
are sampled by $\sim 300$\,AU spacing over the entire field-of-view.
Solid lines denote intensity profiles, and dashed lines are optical thickness
on the same line of sight.
Line profiles in double-horn shape are almost ubiquitously seen in this view with varying
relative ratio of red and blue peaks.}
\label{fig:lprofile}
\end{figure}
\end{center}

\begin{center}
\begin{figure}[htbp]
\epsscale{0.8}
\plotone{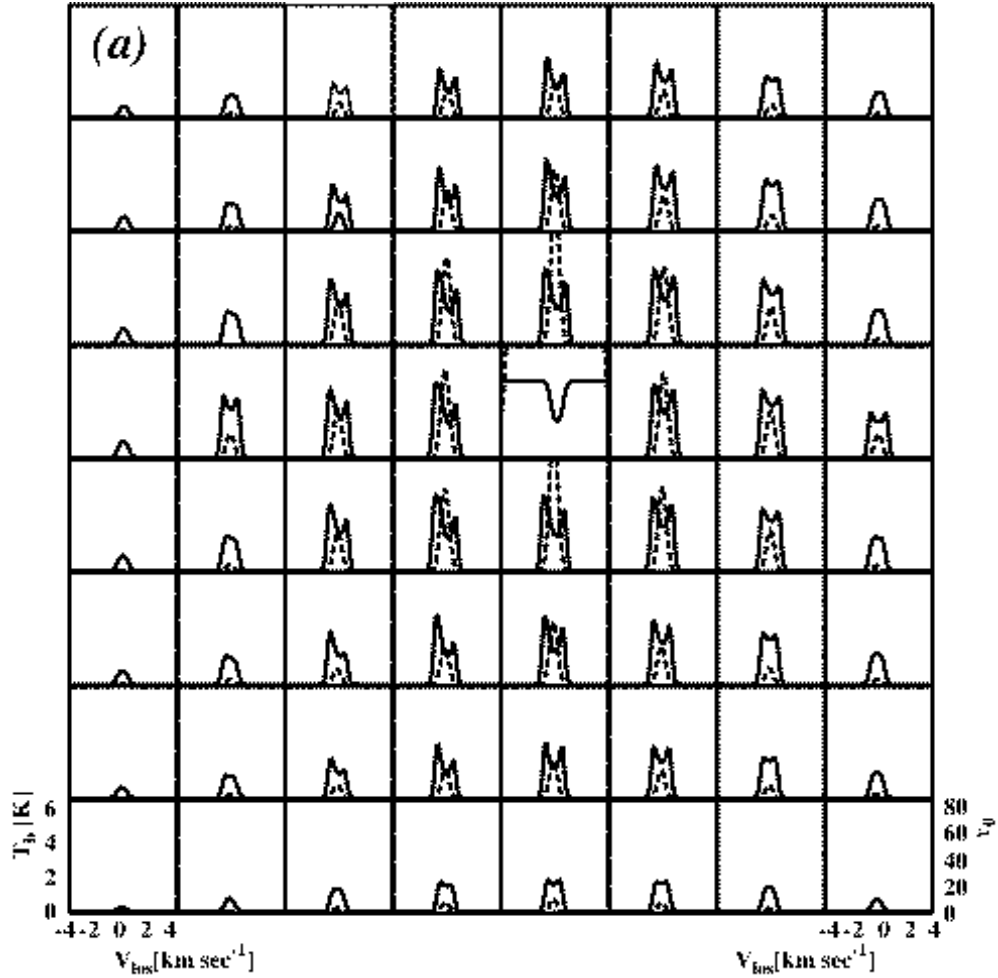} 
\caption{The same distribution maps of line and optical thickness profiles 
as Figure \ref{fig:lprofile}, but for 
the experimental results for case A (panel (a): radial velocity only) and B 
(panel (b): $z$-component of $\boldsymbol{v}_\mathrm{out}$ only). See text for details. 
Panel (a) shows the blue-skewed double-horn profiles almost in the entire field-of-view, 
and panel (b) represents symmetric double-horn profiles in the outflow region.}
\label{fig:lprofile_vr}
\end{figure}
\end{center}

4) In the adopted model in this article, the outflows are driven by a magneto-centrifugal
force \citep{machida2006,machida2008}.
One of the characteristics in magneto-centrifugal force driven flows is their rotation around
the axis.
Its signature can appear in intensity-weighted velocity maps and velocity channel
maps of the synthesized radiation, and should be observable with the ALMA telescope
with a reasonable time ($\sim$ 14 hrs for SiO$(7-6)$ line and $\sim$ 2.6 hrs for 
continuum at 350 GHz). 
Several models have been proposed for launching mechanism of molecular 
outflows \citep[see e.g.,][]{arce2007}, but are still in dispute.
Our results showed a clue to identify the signature of magneto-centrifugal force
driven models (rotation of the outflows: Figure\,\ref{fig:channel1}), and will be an 
important step towards 
understanding of the early stage of star formation.

Based on the limitation of the outer boundary for the radiative transfer calculations, our results 
should be taken as rather qualitative, and not quantitative for comparison to real 
observations.
Further studies will be needed for a confronting theoretical model prediction 
for the ALMA telescope and for revealing, in particular,  
the connection between velocity and emission lines.

\acknowledgments

We thank T. Matsumoto and T. Hanawa for contribution to the nested grid code 
used in hydrodynamic simulation.
M.Y. thanks for intriguing and inspiring discussion and comments of N. Hirano, 
C.-F. Lee,  N. Ohashi and M. Momose, and R. Taam for his critical reading of this  
article.
Numerical computations were partly carried out on VPP5000 and Cray XT4 at the 
Center for Computational Astrophysics (CfCA), National Astronomical
Observatory of Japan, and this research was supported in part by
Grants-in-Aid by the Ministry of Education, Science, and Culture of
Japan (17340059, 16204012,18740104).

\clearpage

%
\begin{table}[htdp]
\caption{Mean optical thickness of dust emission averaged over a field-of-view
$\langle \tau_\nu\rangle$ and the maximum optical thickness in the 
field-of-view $\tau_{\nu,\mathrm{max}}$ for $\theta=$\,30\degr\, view.
Values in this table do not strongly depend on $\theta$ (see Figure\,\ref{fig:cont_sed}).
Three cases with different power-law indices of opacity $\beta=$\, 2, 1.5, and 1 
(Eq.[\ref{eq:kappa}]) are listed.}
\begin{center}
\begin{tabular}{ccccccc}
 & $\nu$ [GHz] & 150 & 220  & 350 & 650 & 850   \cr \hline\hline
$\beta=2.0$ & $\tau_{\nu,\mathrm{max}}$  & 0.37 & 0.79 & 2.01 & 6.94 & 11.9 \cr 
 & $\langle \tau_\nu \rangle$ & 4.31$\times 10^{-4}$ & 9.28 $\times 10^{-4}$ & 2.34 $\times 10^{-3}$ & 8.10 $\times 10^{-3}$ & 0.0138  \cr\hline
 $\beta=1.5$ & $\tau_{\nu,\mathrm{max}}$ & 1.04 & 1.86 & 3.72 & 9.42 & 14.1 \cr
  & $\langle \tau_\nu \rangle$ & 1.22 $\times 10^{-3}$ & 2.17 $\times 10^{-3}$ & 4.35 $\times 10^{-3}$ & 0.0110 & 0.0164  \cr\hline
 $\beta=1.0$ & $\tau_{\nu,\mathrm{max}}$ & 2.95 & 4.33 & 6.89 & 12.8 & 16.7 \cr
 & $\langle \tau_\nu \rangle$ & 3.45 $\times 10^{-3}$ & 5.06 $\times 10^{-3}$ & 8.04 $\times 10^{-3}$ & 0.0149 & 0.0195 \cr
\end{tabular}
\end{center}
\label{tbl:maxtau}
\end{table}%

\clearpage 


%
%

%

\clearpage

%
%

%
%

%
%

%
%

%

%

%
%

%
%

%
%

%
%

%
%

%
%

%
%

%

%

\begin{center}
\begin{figure}[htbp]
\epsscale{0.8}
\figurenum{\ref{fig:lprofile_vr} cont}
\plotone{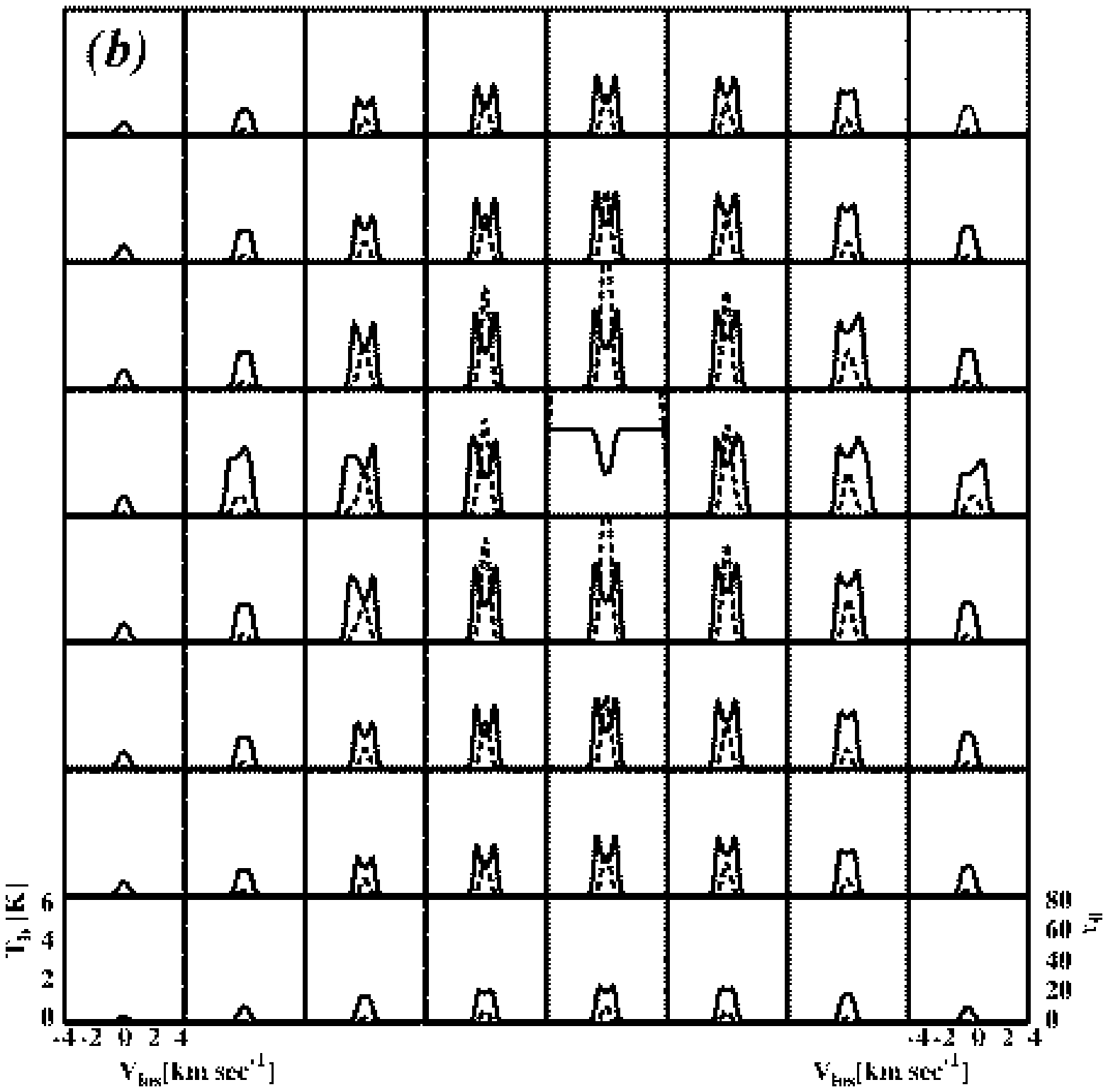}
\caption{Line and optical thickness profile distribution for the case B experimenet (see text). }
\end{figure}
\end{center}

\clearpage

%
\begin{center}
\begin{figure}[htbp]
\plotone{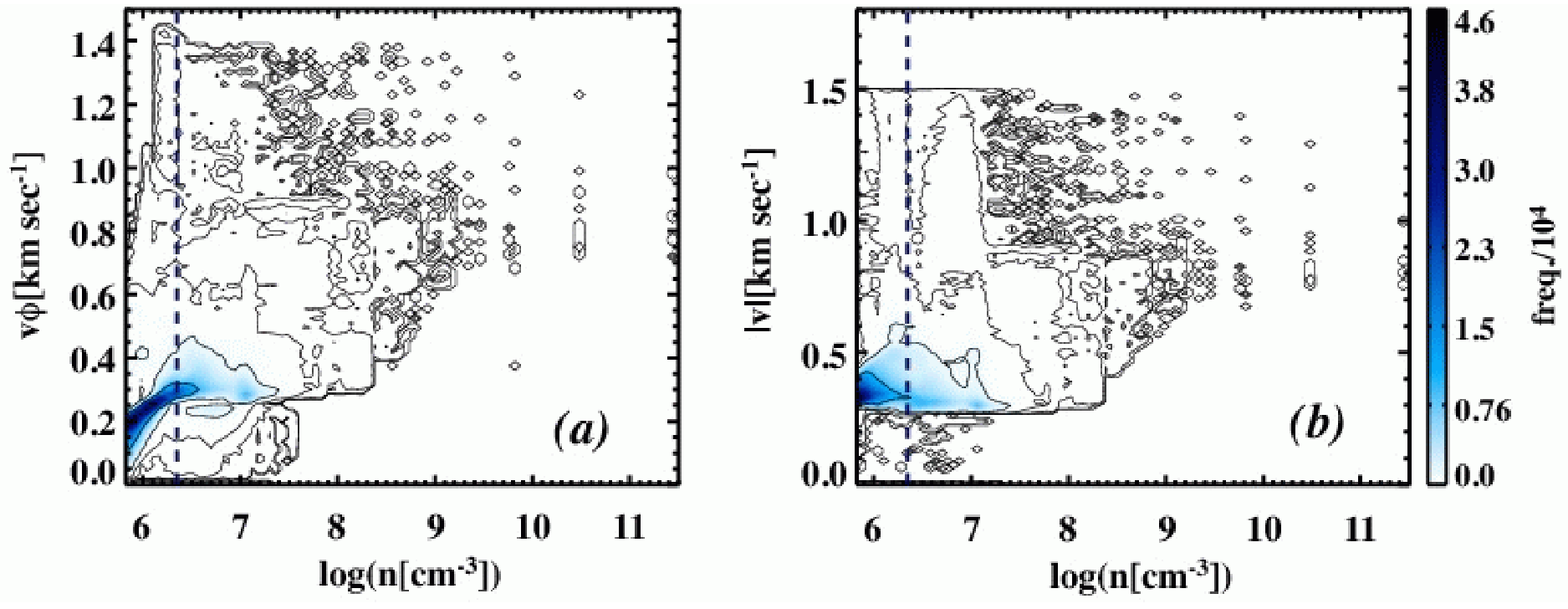}
\caption{Velocity structure of the adopted snapshot data in two-dimensional 
PDF. Panel (a) shows the relation between $v_\phi$ 
and density, and panel (b) shows that of $|\boldsymbol{v}|$ and density in both 
color scale and contours ($0, 10^0, 10^1, 10^2, 10^3$, and $10^4$).}
\label{fig:vel}
\end{figure}
\end{center}

\nobreak

%
\begin{center}
\begin{figure}[bhtp]
\plotone{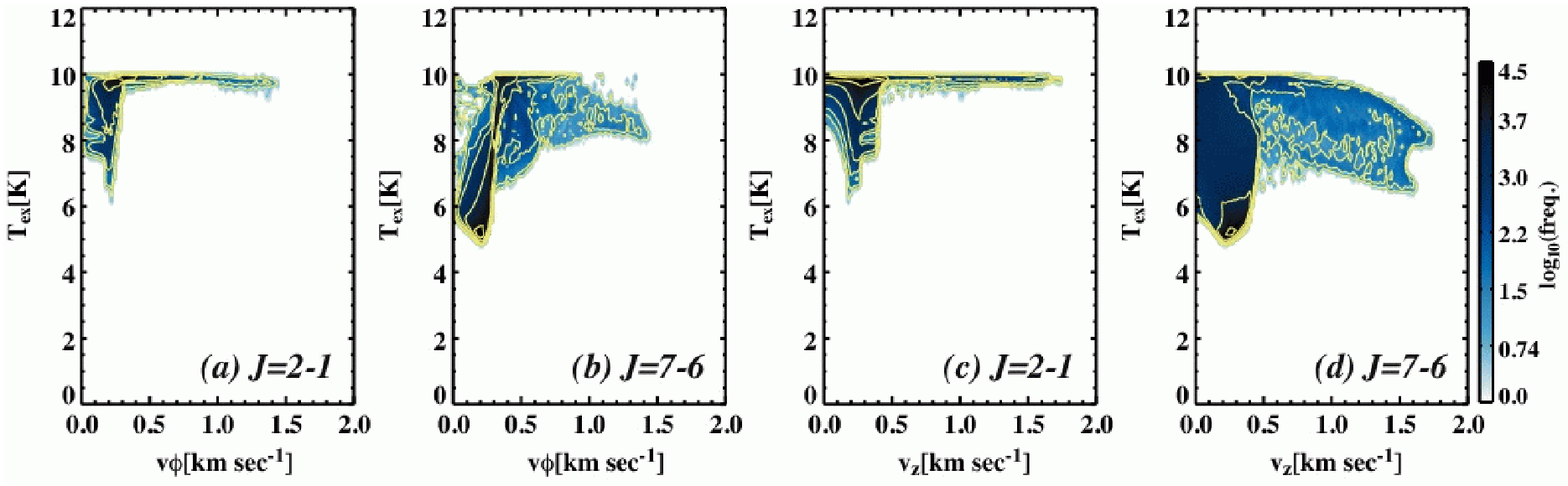}
\caption{Relations between bulk velocity structure of the snapshot and excitation
temperatures of SiO$(2-1)$ and $(7-6)$ lines are shown in two-dimensional PDFs
(solid contours are 0.0, 0.75, 1.5, 2.25, 3.0, 3.75).
Panels (a) and (b) describe the relation of $T_\mathrm{ex}$ and $v_\phi$, and 
panels (c) and (d) describe that of $T_\mathrm{ex}$ and $v_z$.
Higher $J$ line shows a broader scatter for either of $v_\phi$ and $v_z$, 
and has a higher possibility of self-absorption compared to lower $J$ line.}
\label{fig:v_Tex}
\end{figure}
\end{center}

\end{document}